\documentclass[aps, pra, reprint]{revtex4-1}

\usepackage[utf8]{inputenc}
\usepackage[english]{babel}
\usepackage[colorlinks, urlcolor=blue, citecolor=blue]{hyperref}
\usepackage{graphicx}
\usepackage{amssymb, amsmath}
\usepackage{siunitx}

\newcommand{\Hc}{\ensuremath{\mathrm{H.c.}}}

\newcommand{\avg}[1]{\ensuremath{\langle #1\rangle}}
\newcommand{\Hint}{\ensuremath{H_\mathrm{int}}}
\newcommand{\inpt}{\ensuremath{\mathrm{in}}}

\newcommand{\eff}{\ensuremath{\mathrm{eff}}}
\newcommand{\vect}[1]{\ensuremath{\mathbf{#1}}}
\newcommand{\nbar}{\ensuremath{\bar{n}}}
\newcommand{\Vsq}{\ensuremath{V_\mathrm{sq}}}
\newcommand{\Vasq}{\ensuremath{V_\mathrm{asq}}}

\begin{document}

\title{Strong mechanical squeezing for a levitated particle by coherent scattering}

\author{Ond\v{r}ej \v{C}ernot\'ik}
\email{ondrej.cernotik@upol.cz}
\affiliation{Department of Optics, Palack\'y University, 17. listopadu 1192/12, 77146 Olomouc, Czechia}

\author{Radim Filip}
\email{filip@optics.upol.cz}
\affiliation{Department of Optics, Palack\'y University, 17. listopadu 1192/12, 77146 Olomouc, Czechia}

\date{\today}

\begin{abstract}
	Levitated particles are a promising platform for precision sensing of external perturbations and probing the boundary between quantum and classical worlds.
	A critical obstacle for these applications is the difficulty of generating nonclassical states of motion which have not been realized so far.
	Here, we show that strong squeezing of the motion of a levitated particle below the vacuum level is feasible with available experimental parameters.
	Using suitable modulation of the trapping potential (which is impossible with clamped mechanical resonators) and coherent scattering of trapping photons into a cavity mode, we explore several strategies to achieve strong phase-sensitive suppression of mechanical fluctuations.
	We analyze mechanical squeezing in both transient and steady-state regimes, and discuss conditions for preparing nonclassical mechanical squeezing.
	Our results pave the way to full, deterministic optomechanical control of levitated particles in the quantum regime.
\end{abstract}

\maketitle

\section{Introduction}

Cavity optomechanics~\cite{Aspelmeyer2014}, in which optical fields interact with mechanical elements via radiation pressure, has a tremendous potential for sensing of weak forces~\cite{Forstner2012,Buchmann2016,DeLepinay2017,Ockeloen-Korppi2018a} and testing fundamental physical theories~\cite{Pikovski2012,Pfister2016}.
Particularly levitated nanoparticles~\cite{Chang2009,Romero-Isart2011,Millen2019} represent---owing to lack of clamping losses---an interesting platform for metrology~\cite{Prat-Camps2017,Hebestreit2018,Blakemore2019}, thermodynamics~\cite{Gieseler2018,Siler2018}, and probing the quantum--classical boundary~\cite{Romero-Isart2011a,Bateman2014} or other fundamental theories~\cite{Moore2014,Rider2016}.
Experimental techniques for cooling~\cite{Gieseler2012,Kiesel2013,Asenbaum2013,Conangla2019,Tebbenjohanns2019} and thermal squeezing~\cite{Rashid2016} of their center-of-mass motion, as well as for controlling their rotations~\cite{Ahn2018,Reimann2018} and libration~\cite{Hoang2016,Kuhn2017}, have been firmly established.
Despite these efforts and results, genuinely nonclassical states of motion of levitated particles remain elusive.

Here, we propose techniques for achieving mechanical squeezing with levitated particles and show that quantum squeezing (i.e., squeezing below the vacuum level) is feasible with state-of-the-art systems using parametric and dissipative squeezing.
Parametric squeezing has been studied for clamped mechanical oscillators~\cite{Mari2009,Liao2011,Chowdhury2019} for which, however, only the optical spring can be modulated using a suitable driving field.
For levitated particles, direct modulation of the trapping potential is possible using an amplitude-modulated trapping field, resulting in particularly strong squeezing in the transient regime.
Dissipative squeezing is also well known in optomechanics~\cite{Kronwald2013,Pirkkalainen2015,Wollman2015,Lei2016};
we show here that adding parametric to dissipative squeezing (by modulating the trapping potential) enables stronger suppression of mechanical fluctuations than either technique alone.
Crucially, our strategies are deterministic (obviating the need for efficient measurements typical for conditional preparation of squeezing~\cite{Clerk2008,Genoni2015,Rakhubovsky2019}) and do not require nonlinear interactions~\cite{Nunnenkamp2010,Asjad2014,Lu2015}.

In our proposals, we employ coherent scattering of the trapping beam into an empty cavity mode~\cite{Vuletic2001,Gonzalez-Ballestero2019} instead of the usual dispersive optomechanical interaction.
This technique has, so far, been used to cool the motion of trapped ions~\cite{Leibrandt2009} and, recently, levitated particles~\cite{Delic2019,Windey2019} (including cooling from the room temperature to the quantum ground state~\cite{Delic2019b});
we show that it can be used for more advanced control of mechanical motion.
In our case, amplitude modulation of the trapping field results in modulation of both the trapping potential and optical spring, resulting in strong parametric oscillations of motion with a low instability threshold, allowing  strong mechanical squeezing to be generated.
Furthermore, as coherent scattering enables all mechanical modes to be coupled to the same cavity mode, our results can be generalized to multiple dimensions, serving as a first step towards preparing complex nonclassical states of the motion of levitated particles.

\section{Theoretical model}

\begin{figure} 
	\centering
	\includegraphics[width=\linewidth]{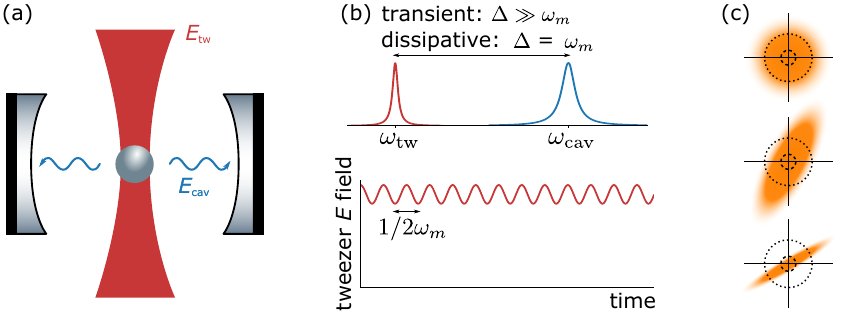}
	\caption{\label{fig:Scheme}
	Generating mechanical squeezing by coherent scattering.
	(a) Experimental setup, in which photons from the trapping beam are scattered off of a nanoparticle into an empty cavity mode.
	(b) Tools for creating mechanical squeezing:
	A cavity field with a suitable detuning from the tweezer (top) and amplitude modulation of the trapping field at twice the mechanical frequency (bottom).
	(c) Starting from an initial thermal state (Wigner function shown on top), one can create a thermal squeezed state (middle) or even a quantum squeezed state (bottom).
	The large dotted circle represents the variance of the initial thermal state, the smaller dashed circle shows the shot noise level.}
\end{figure}

The system we consider is depicted in Fig.~\ref{fig:Scheme}(a).
A nanoparticle is trapped in an optical tweezer and placed in a cavity; scattering of tweezer photons off of the particle into the cavity mode gives rise to optomechanical interaction as described by Gonzalez-Ballestero et al.~\cite{Gonzalez-Ballestero2019}.
The total Hamiltonian of such a system is given by
\begin{equation}
    H = H_\mathrm{part} + H_\mathrm{field} + \Hint.
\end{equation}
The first term, $H_\mathrm{part} = \vect{P}^2/2m$, describes the free Hamiltonian of a particle with momentum $\vect{P}$ and mass $m$.
Second, $H_\mathrm{field} = \omega_\mathrm{cav} c^\dagger c$ accounts for the cavity field (annihilation operator $c$, angular frequency $\omega_\mathrm{cav}$).
Finally, $\Hint = -\frac{1}{2}\alpha_p \vect{E}^2(\vect{R},t)$ characterizes the interaction between the fields and the particle motion;
here $\alpha_p$ is the nanoparticle polarizability and $\vect{E}(\vect{R},t)$ the electric field at the particle position $\vect{R}$~\cite{Romero-Isart2011}.

To evaluate the interaction Hamiltonian,
we decompose the electric field into the tweezer and cavity components, $\vect{E}(\vect{R}) = \vect{E}_\mathrm{cav}(\vect{R}) + \vect{E}_\mathrm{tw}(\vect{R})$.
For a Gaussian trapping beam and small displacements, the square of the tweezer field gives the harmonic potential $V = \sum_j\frac{1}{2}m\omega_j^2R_j^2$ with frequencies determined by the strength of the field, density and polarizability of the particle, as well as geometric properties (waist and Rayleigh range) of the tweezer~\cite{Romero-Isart2011}.
The square of the cavity field gives, in a similar manner, the conventional dispersive coupling between the particle and the cavity field ($\hbar=1$ in the following)~\cite{Romero-Isart2011}
\begin{equation}
	H_\mathrm{om} = - g_0(X\cos\theta+Y\sin\theta)c^\dagger c.
\end{equation}
Here, $X$ and $Y$ describe the displacements along and perpendicular to the tweezer polarization (they are both perpendicular to the tweezer axis) and $\theta$ is the angle between the tweezer and cavity polarizations.

Finally, the product of the tweezer and cavity fields describes coherent scattering~\cite{Gonzalez-Ballestero2019},
\begin{align}
\begin{split}
    H_\mathrm{sc} &= -\lambda_0\cos\phi(c+c^\dagger) -i\lambda_zZ\cos\phi(c-c^\dagger)\\
    	&\quad - (\lambda_xX\cos\theta+\lambda_yY\sin\theta)\sin\phi(c+c^\dagger),
\end{split}
\end{align}
where $\phi$ specifies the position of the particle within the standing wave of the cavity.
Particularly, cavity nodes are characterized by $\cos\phi = 0$ and antinodes by $\cos^2\phi = 1$.
The relative strength of the interactions $\lambda_{x,y,z}$ can thus be tuned by controlling the particle position and polarization of the tweezer field.
Using parallel tweezer and cavity polarizations ($\cos\theta = 1$) and placing the particle at a node of the cavity field ($\sin\phi = 1$), we can cancel coupling to the $Y$ and $Z$ modes.
The total Hamiltonian describing the interaction between the particle motion (along $X$ only) and the cavity field is then given by
\begin{equation}\label{eq:HBasic}
	H = \Delta c^\dagger c+\frac{\omega_x}{2}(x^2+p^2) - \lambda (c+c^\dagger)x;
\end{equation}
note that we work with the dimensionless position and momentum quadratures $x = X/x_\mathrm{zpf}$, $p = P_x/p_{x,\mathrm{zpf}}$ with the commutator $[x,p]=i$.
The dispersive optomechanical interaction vanishes at a cavity node, and we describe the cavity mode in a frame rotating at the tweezer frequency $\omega_\mathrm{tw}$ with the detuning $\Delta = \omega_\mathrm{cav}-\omega_\mathrm{tw}$;
we also defined $\lambda = \lambda_x$.

To account for decoherence effects, we describe the dynamics using Langevin equations
\begin{subequations}\label{eq:Langevin}
\begin{align}
	\dot{c} &= -(\kappa+i\Delta)c +i\lambda x +\sqrt{2\kappa}c_\inpt,\label{eq:LangevinCav}\\
	\dot{x} &= \omega_x p,\\
	\dot{p} &= -\omega_x x-\gamma p+\lambda(c+c^\dagger) +\xi
\end{align}
\end{subequations}
with cavity decay rate $\kappa$ and mechanical damping rate $\gamma$.
The stochastic noise operator $c_\inpt$ is the usual delta-correlated cavity input vacuum noise, $\avg{c_\inpt(t)c_\inpt^\dagger(t')} = \delta(t-t')$, and the thermal noise operator $\xi$ satisfies $\avg{\xi(t)\xi(t')} = 2\gamma(2\nbar+1)\delta(t-t')$ with $\nbar$ characterizing the average thermal occupation of the mechanical bath.

\section{Squeezing techniques}

\subsection{Adiabatic elimination of cavity dynamics}\label{ssec:elimination}

As a first step, we assume a simple scenario where the cavity detuning is large compared to other system parameters.
In this regime, the cavity remains unpopulated and only indirectly affects the mechanical motion;
it can thus be adiabatically eliminated from the dynamics.
Formal integration of Eq.~\eqref{eq:LangevinCav} yields
\begin{equation}
	c(t) = \frac{1}{\kappa+i\Delta}(i\lambda x+\sqrt{2\kappa}c_\inpt).
\end{equation}
Plugging this expression back into the equations of motion for the particle gives
\begin{subequations}\label{eq:EOMelimination}
\begin{align}
	\dot{x} &= (\omega_\eff+\zeta_\eff)p,\\
	\dot{p} &= -(\omega_\eff-\zeta_\eff)x -\gamma p +\xi +\frac{2\lambda\sqrt{\kappa}}{\kappa^2+\Delta^2}(\kappa X_\inpt +\Delta Y_\inpt)\label{eq:PElimination}
\end{align}
\end{subequations}
with the effective oscillation frequency $\omega_\eff = \omega_x - \Delta\lambda^2/(\kappa^2+\Delta^2)$ and squeezing parameter $\zeta_\eff = \Delta\lambda^2/(\kappa^2+\Delta^2)$.
The interaction with a far-detuned cavity field thus turns the particle into a parametric oscillator.
The cavity also contributes with two quadrature noise terms defined as $X_\inpt = (c_\inpt+c_\inpt^\dagger)/\sqrt{2}$, $Y_\inpt = -i(c_\inpt-c_\inpt^\dagger)/\sqrt{2}$.
In the limit of large detuning, $\Delta\gg\kappa$, the coupling to the cavity input noise can be reduced and approximated by the term $(2\lambda\sqrt{\kappa}/\Delta) \times Y_\inpt$;
moreover, $\omega_\eff\approx \omega_x-\lambda^2/\Delta$, $\zeta_\eff\approx \lambda^2/\Delta$.

To find the resulting squeezing, we solve the differential Lyapunov equation for the covariance matrix of the mechanical mode as described in Appendix~\ref{app:numerics}.
Diagonalization of the covariance matrix reveals the variance of the squeezed and antisqueezed quadratures, denoted by $\Vsq,\Vasq$.
We then quantify the noise distribution by the ratio of the two quadratures (which we call the squeezing degree in the following), $\eta = \Vsq/\Vasq$, so that presence of squeezing corresponds to $\eta < 1$.
To further distinguish between classical and quantum squeezing, we are also interested in the value of the squeezed variance $\Vsq$;
value below the vacuum level, $\Vsq < 1$, implies the generation of a nonclassical squeezed state.

\begin{figure} 
	\centering
	\includegraphics[width=\linewidth]{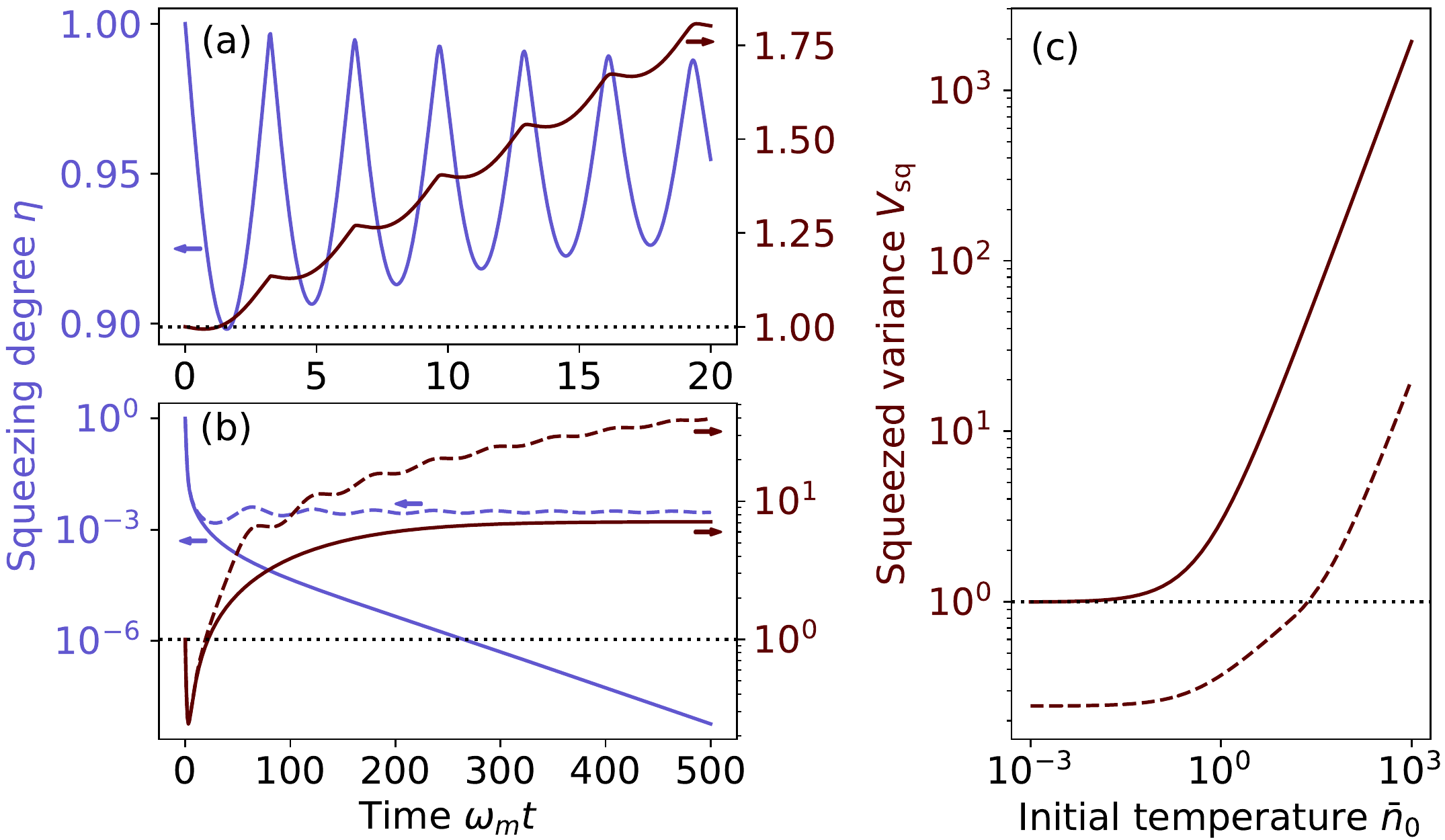}
	\caption{\label{fig:Transient}
	Mechanical squeezing generated with a detuned cavity.
	(a) Squeezing degree $\eta$ (light blue) and squeezed variance $\Vsq$ (dark red) versus time for the parameters $\lambda/\omega_x = 0.3$, $\kappa/\omega_x = 0.2$, $Q_m = \omega_x/\gamma = 10^9$, $\nbar = 2\times 10^7$, $\Delta/\omega_x = 5$.
	These values are similar to the parameters in the recent experiment on ground state cooling by coherent scattering~\cite{Delic2019b} except for a slightly larger coupling (possible with a larger particle), smaller thermal decoherence (achievable by operating at a higher vacuum), and smaller cavity decay rate (which is necessary to limit cavity input noise);
	the thermal noise level corresponds to a \SI{300}{\kilo\hertz} mechanical mode at a temperature of \SI{300}{\kelvin}.
	(b) Squeezing plotted for coupling at the instability threshold $\lambda_\mathrm{th}/\omega_x\approx 1.5825$ (solid lines);
	the dashed lines show the squeezing degree and squeezed variance for coupling just below the threshold, $\lambda/\omega_x = 1.58$.
	(c) Squeezed variance optimized over time as a function of the initial temperature of the mechanical mode for the parameters from panel (a) (solid line) and at the instability threshold (dashed).
	The initial mechanical state is the thermal state with variance $V_0 = 2\nbar_0 + 1$;
	in panels (a,b), we assume precooling to the mechanical ground state, $\nbar_0 = 0$.
	The horizontal dotted lines in all panels show the vacuum variance.}
\end{figure}

We plot the resulting squeezing versus time in Fig.~\ref{fig:Transient} (a);
the system parameters are similar to the recent demonstration of ground state cooling via coherent scattering~\cite{Delic2019b} (see figure caption for details).
Nonclassical squeezing puts extremely stringent conditions on the system parameters, requiring precooling the mechanical motion to its quantum ground state and stronger coupling than is currently available.
Stronger squeezing is generally possible with a larger coupling rate; the maximum is reached at the onset of dynamical instability [Fig.~\ref{fig:Transient}(b)].
This occurs when the first term on the right-hand side of Eq.~\eqref{eq:PElimination} vanishes, which is achieved for $\lambda_\mathrm{th} = \sqrt{\omega_x(\kappa^2+\Delta^2)/2\Delta}$.
Below threshold, the squeezed variance (achieved in the middle of the first oscillation period) decreases with growing coupling, reaching its minimum at the onset of instability.
At later times, the variance grows with time below threshold;
above threshold, it reaches a quasistationary value.
(Note that the squeezing degree is gradually decreasing, implying that the antisqueezed variance diverges which causes the particle to escape the trap.)
The optimum squeezing as a function of the initial mechanical occupation is shown in Fig.~\ref{fig:Transient}(c).
Although quantum squeezing is generally possible, realistic experimental parameters (particularly the thermal decoherence rate) allow only for thermal squeezing.

\subsection{Trapping field modulation}\label{ssec:modulation}

Strong parametric squeezing can be achieved when the mechanical spring constant is modulated at twice the mechanical frequency.
This situation can be achieved when the amplitude of the trapping beam, $E_0$ (assumed constant so far), is modulated as $E_0(t) = E_0[1+\alpha\cos(2\omega_x t+\phi)]$ with depth $\alpha\in (0,1)$ and phase $\phi$.
Since the trapping potential is given by the square of the tweezer field, $\omega_x\propto E_0^2(t)$, and the optomechanical interaction is linear in the tweezer field, $\lambda\propto E_0(t)$, the modulation modifies the Hamiltonian according to
\begin{align}\label{eq:SMModulated}
    H &= \frac{\omega_x}{2}p^2+\frac{\omega_x}{2}[1+\alpha\cos(2\omega_x t+\phi)]^2x^2 + \Delta c^\dagger c\nonumber\\
    &\quad -\lambda[1+\alpha\cos(2\omega_x t+\phi)]x(c+c^\dagger).
\end{align}
In the rotating frame with respect to the free mechanical oscillations $H_m = \frac{1}{2}\omega_x(x^2+p^2)$ and under the rotating wave approximation, the mechanical potential gives rise to the parametric oscillations $H_\mathrm{par} = \frac{1}{4}\omega_x\alpha(\alpha b^\dagger b + b^2e^{i\phi} +b^{\dagger 2}e^{-i\phi})$,
the strength of which is fully controlled by the modulation depth $\alpha$.

To account also for the modulated optomechanical coupling, we adiabatically eliminate the cavity field from the dynamics.
Using the approach outlined in Appendix~\ref{app:modulation}, we obtain the effective Hamiltonian
\begin{equation}
	H_\eff = \frac{\omega_\eff}{2}(x^2+p^2) +\frac{\zeta_\eff}{2}[\cos\phi(x^2-p^2) -\sin\phi(xp+px)]
\end{equation}
with effective frequency $\omega_\eff = [\omega_x\Delta\alpha^2-2\lambda^2(\alpha+\alpha^2)]/4\Delta$ and squeezing parameter $\zeta_\eff = \alpha(\omega_x\Delta -2\lambda^2)/2\Delta$.
The optical input noise can be included in the dynamics via the effective Langevin equations
\begin{subequations}\label{eq:LangEffective}
\begin{align}\label{eq:LangEffectiveX}
	\dot{x} &= \omega_\eff p-\zeta_\eff(x\sin\phi+p\cos\phi)+\frac{\lambda\sqrt{\kappa}}{\Delta}X_\inpt,\\
	\dot{p} &= -\omega_\eff x-\zeta_\eff(x\cos\phi-p\sin\phi)-\gamma p+\xi+\frac{\lambda\sqrt{\kappa}}{\Delta}Y_\inpt.
\end{align}
\end{subequations}
Formally, the dynamics are similar to the previous case with two important distinctions:
First, the modulation phase $\phi$ allows us to choose which quadrature to squeeze, permitting squeezing in a quadrature which is minimally coupled to external noise sources.
Second, parametric modulation of the mechanical frequency allows, owing to the reduced effective mechanical frequency, for stronger squeezing to be observed;
note that we operate in a frame rotating at the mechanical frequency so $\omega_\eff$ is given only by the self-energy arising from the optomechanical interaction and parametric driving.

\begin{figure} 
	\centering
	\includegraphics[width=\linewidth]{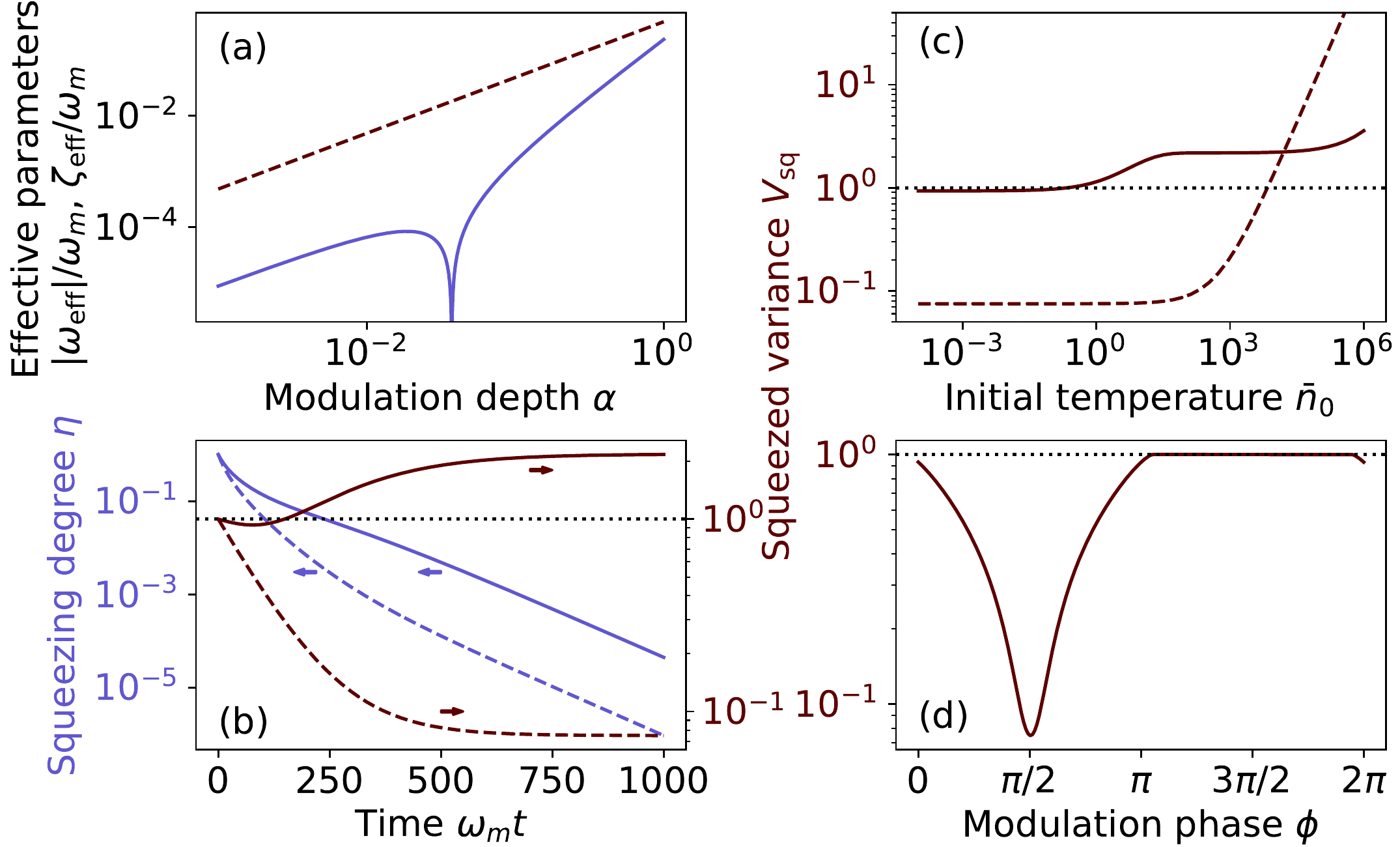}
	\caption{\label{fig:Parametric}Mechanical squeezing with tweezer modulation.
	(a) Effective mechanical frequency $|\omega_\eff|$ (solid blue line) and squeezing $\zeta_\eff$ (dashed red line) versus modulation depth $\alpha$.
	For small modulation depths, the effective frequency is negative, $\omega_\eff < 0$, and becomes positive for $\alpha\gtrsim 0.037$.
	(b) Time-dependence of the squeezing degree $\eta$ (light blue) and squeezed variance $\Vsq$ (dark red) for modulation depth $\alpha = 0.01$ and phase $\phi = 0$ (solid) or $\phi = \pi/2$ (dashed).
	(c,d) Squeezed variance optimized over time as a function of the initial mechanical occupation $\nbar_0$ (c) and modulation phase (d);
	the solid and dashed lines have the same meaning as in panel (b).
	System parameters are the same as in Fig.~\ref{fig:Transient}(a) with $\nbar_0 = 0$ for panels (b,d).}
\end{figure}

We examine the resulting squeezing in Fig.~\ref{fig:Parametric}.
A comparison of the effective mechanical frequency $\omega_\eff$ and squeezing $\zeta_\eff$ [panel (a)] reveals the advantage of using tweezer modulation:
for the chosen parameters, the dynamics are always unstable (since $|\omega_\eff|<\zeta_\eff$), allowing strong squeezing as shown in panel (b).
The optimum squeezing is further analyzed in panels (c,d) as a function of the initial mechanical occupation $\nbar_0$ and the modulation phase $\phi$.
Strong quantum squeezing is possible for a broad range of initial temperatures in the quasistationary regime where strong suppression of quantum fluctuations occurs.
The optimal choice of modulation phase is $\phi = \pi/2$ for which the amplitude quadrature is squeezed (with $\omega_\eff = 0$ or, approximately, with $|\omega_\eff|\ll\zeta_\eff$);
this result is expectable since the amplitude quadrature is decoupled from the thermal mechanical bath (which was one of the main limiting factors in the previous case) and cavity noise affects both quadratures equally.

\subsection{Adding dissipative squeezing}\label{ssec:dissipative}

Additional possibility is to tune the cavity close to resonance with the tweezer, allowing more photons to be scattered.
When we choose the detuning $\Delta = \omega_x$ and work with a sideband resolved system (such that $\kappa\ll\omega_x$), the overall Hamiltonian becomes [in rotating frame with respect to $H_0 =\frac{1}{2}\omega_x(x^2+p^2)+ \Delta c^\dagger c$; see Appendix~\ref{app:dissipative}]
\begin{equation}\label{eq:HintModulated}
	\Hint = \frac{\omega_x\alpha}{4}(\beta^2+\beta^{\dagger 2}-\alpha\beta^\dagger\beta) -\lambda_\eff(\beta c^\dagger +\beta^\dagger c).
\end{equation}
Here, we introduced the mechanical Bogoliubov mode $\beta = (2b+\alpha b^\dagger)/\sqrt{4-\alpha^2}$ (with $\phi = 0$ for simplicity; our numerical simulations show that the resulting squeezing is independent of the modulation phase as we show in Appendix~\ref{app:dissipative}) and the effective coupling rate $\lambda_\eff = \lambda\sqrt{(4-\alpha^2)/8}$.
The optomechanical interaction [the second term in Eq.~\eqref{eq:HintModulated}] thus cools the Bogoliubov mode $\beta$ to its ground state~\cite{Kronwald2013}, resulting, unlike the previous two methods, in steady-state mechanical squeezing.

Although dissipative engineering can generally be used to generate steady-state squeezing below the \SI{3}{\deci\bel} limit of parametric squeezing, its performance here is limited by admissible modulation depths.
The squeezed variance for the Bogoliubov mode $\beta$ in its ground state depends on the modulation depth via $V_\alpha = (2-\alpha)/(2+\alpha)$ which reaches the value of $\frac{1}{3}$ for $\alpha\to 1$;
this limit corresponds to about \SI{4.8}{\deci\bel} of squeezing.
The generated squeezing can, however, be further improved by parametrically squeezing the Bogoliubov mode as indicated by the first term in Eq.~\eqref{eq:HintModulated}, allowing this limit (as well as the \SI{3}{\deci\bel} limit of steady-state parametric squeezing) to be surpassed.

\begin{figure} 
	\centering
	\includegraphics[width=\linewidth]{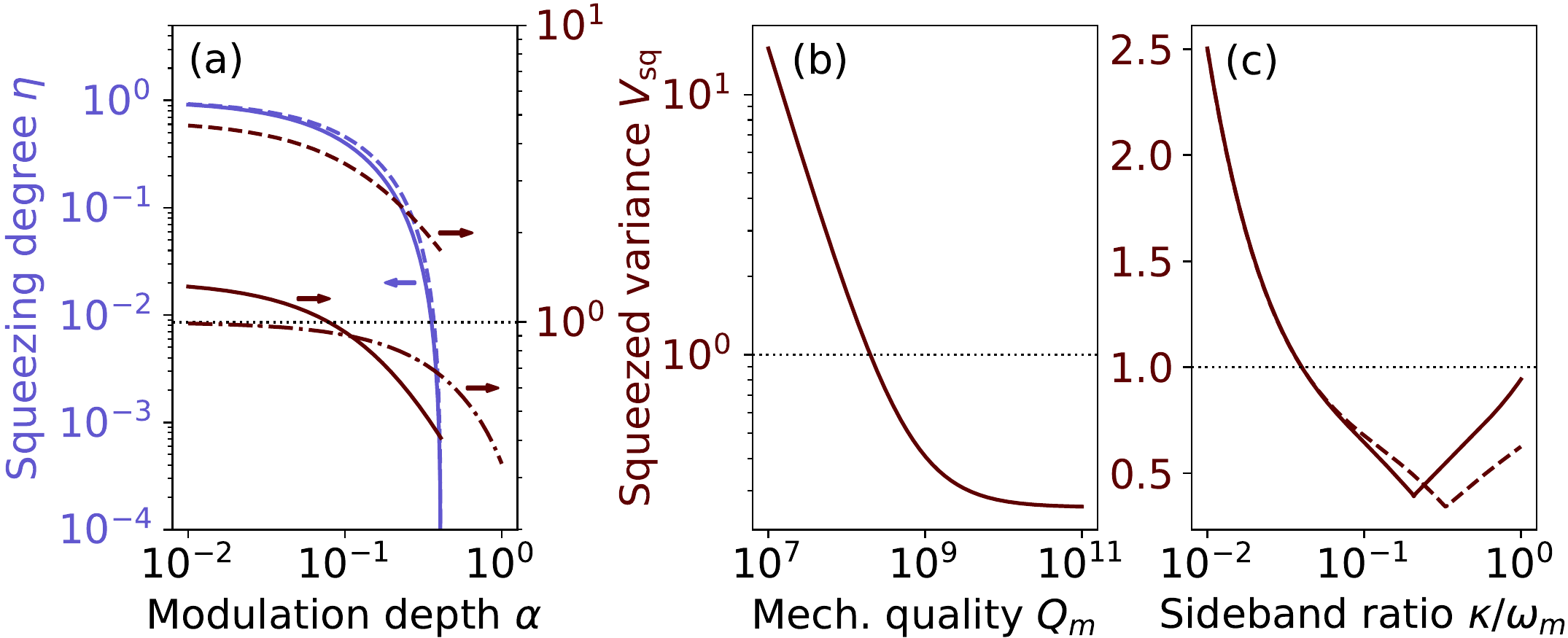}
	\caption{\label{fig:Modulation}Dissipative and parametric mechanical squeezing.
	(a) Squeezing degree $\eta$ (light blue) and squeezed variance $V_\mathrm{sq}$ (dark red) versus modulation depth.
	Solid lines show results for $Q_m = 10^9$, dashed curves are for $Q_m = 10^8$;
	other parameters are the same as in previous figures.
	The dot-dashed line shows the squeezed variance $V_\alpha$ obtained by cooling the Bogoliubov mode $\beta$ to its ground state.
	(b) Squeezed variance $V_\mathrm{sq}$ optimized over the modulation depth as a function of the mechanical quality factor $Q_m$.
	(c) The optimized squeezed variance versus the sideband ratio $\kappa/\omega_x$ for $\lambda/\omega_x = 0.3$ (solid) and $\lambda/\omega_x = 0.5$ (dashed).}
\end{figure}

We analyze this steady-state squeezing in Fig.~\ref{fig:Modulation}.
When the modulation depth is increased [panel (a)], the amount of squeezing increases as the system approaches instability; at its onset, the squeezed variance is minimized.
The squeezing degree $\eta$ is largely unaffected by thermal decoherence;
the squeezed variance $V_\mathrm{sq}$ can, for sufficiently high mechanical quality, be smaller than the variance obtained by cooling the Bogoliubov mode $\beta$ to its ground state.
Additionally, the minimum variance shown here $V_\mathrm{sq,min}\approx 0.4 < \frac{1}{2}$, demonstrating that the \SI{3}{\deci\bel} limit, which applies to parametric squeezing in the steady state, can be surpassed as well.
These results thus add to the existing body of work that shows enhanced squeezing when multiple squeezing techniques are combined~\cite{Lei2016,Hu2018}.
One could na\"{i}vely expect that the resulting squeezing will be limited to $\Vsq = \frac{1}{2}V_\alpha$, obtained by addition of the two squeezing effects;
however, for sufficiently low mechanical decoherence, the squeezed variance can reach $\Vsq \approx 0.26$ with the parameters of Fig.~\ref{fig:Modulation} [panel (b)], which is smaller than the simple bound $\frac{1}{2}V_\alpha= \frac{1}{3}$ for $\alpha = 0.4$.

The optimum mechanical squeezing (i.e., squeezing obtained at the onset of instability) is further analyzed in Fig.~\ref{fig:Modulation}(b,c).
The scheme is resilient against thermal noise with thermal decoherence rates up to $\gamma\nbar/\omega_x\approx 0.2$ allowing squeezing below the vacuum level [panel (b)].
Finally, there is a nontrivial dependence between the chosen coupling rate and optimal cavity decay rate [panel (c)];
larger coupling rates require faster cavity decay to achieve optimal cooling performance.
Both coupling rate and cavity decay are limited as we employed the rotating wave approximation to obtain the Hamiltonian~\eqref{eq:HintModulated};
the resulting squeezing can thus be improved primarily by reducing the mechanical decoherence.

\section{Discussion and conclusions}

Further improvements in the generated squeezing are possible using squeezed cavity input.
In the case of the simple parametric oscillator discussed in Sec.~\ref{ssec:elimination}, squeezing of the input quadrature $X_\mathrm{sq,in} = (\kappa X_\inpt +\Delta Y_\inpt)/\sqrt{\kappa^2+\Delta^2}$ reduces the overall noise in the mechanical equations of motion [cf. Eqs.~\eqref{eq:EOMelimination}];
unfortunately, for realistic experimental parameters~\cite{Delic2019b}, the cavity input noise is much weaker than thermal decoherence and input squeezing thus provides no advantage.
The situation is different for modulated trapping field where we have the freedom to choose the squeezed mechanical quadrature.
Squeezing of the amplitude quadrature is limited only by the cavity input noise coupled to this quadrature [see Eq.~\eqref{eq:LangEffectiveX}];
we thus expect that reducing the noise in the input amplitude quadrature reduces the mechanical position variance by the same factor.
Finally, input squeezing might also improve the performance of dissipative squeezing as squeezing at the cavity input can be used to suppress unwanted scattering processes in optomechanics~\cite{Asjad2016,Clark2017};
we leave a detailed analysis of this effect for future work~\cite{Pietikainen2019}.

Similar results could, in principle, be achieved also with conventional dispersive optomechanics using modulation of the tweezer amplitude and two-tone driving of the cavity, but our approach offers several advantages:
Since both the optical potential and optomechanical interaction are derived from the tweezer field, their relative phase---crucial for efficient squeezing---is automatically locked;
additionally, absorption heating of the particle is reduced since a single optical beam provides both trapping and coupling to the cavity field.
The most crucial advantage, however, lies in the prospect of coupling multiple mechanical modes via their interaction with a single cavity mode.
A straightforward generalization of our schemes should allow the creation of two-mode squeezing between two motional modes of the particle;
in this context, we note that both parametric~\cite{Pontin2016} and dissipative~\cite{Ockeloen-Korppi2018} two-mode mechanical squeezing have been realized in optomechanical systems.
In the long term, full quantum control of motion should be possible, first in the Gaussian regime (via linear coupling of all three motional modes to the same cavity mode) and later in the non-Gaussian regime (when nonlinear optomechanical interactions or anharmonic trapping potentials are added~\cite{Neumeier2018,Delic2019a,Rakhubovsky2019a,Setter2019}).

In conclusion, we demonstrated that strong squeezing of motion of levitated particles is possible in state-of-the-art systems.
With a combination of parametric amplification (achievable by modulating the trapping beam) and dissipation (using a cavity mode to cool down a mechanical Bogoliubov mode), squeezing below the vacuum level is possible in the steady state.
Even stronger squeezing, albeit in the transient regime, is possible with parametric amplification alone;
although this might appear as a limitation at first, strong transient squeezing combined with single-phonon control~\cite{Riedinger2016,Chu2018,Satzinger2018} can be used to prepare macroscopic superposition states of macroscopic objects for applications in sensing~\cite{Hebestreit2018} or testing quantum mechanics~\cite{Bateman2014}.
Unlike existing proposals, our unconditional strategy relies on coherent scattering of tweezer photons into the cavity, demonstrating the potential of coherent scattering as a general tool for controlling the motion of levitated particles.
With the prospect of engineering interactions between mechanical modes via coherent scattering, our work thus presents an important step towards full quantum control of motion of levitated particles.

\begin{acknowledgments}
We thank Markus Aspelmeyer, Uro\v{s} Deli\'{c}, and Nikolai Kiesel for fruitful discussions on the feasibility of our proposals.
We gratefully acknowledge support by the project CZ.02.1.01/0.0/0.0/16\textunderscore026/0008460 of MEYS \v{C}R.
R.F. has also been supported by Project 19-17765S of the Czech Science Foundation, national funding from MEYS, and funding from European Union’s Horizon 2020 (2014–2020) research and innovation framework programme under Grant Agreement No. 731473 (project 8C18003 TheBlinQC).
Project TheBlinQC has received funding from the QuantERA ERA-NET Cofund in Quantum Technologies implemented within the European Union’s Horizon 2020 Programme.
\end{acknowledgments}

\appendix

\section{Numerical methods}\label{app:numerics}

To estimate the mechanical squeezing numerically, we find the covariance matrix of the system by solving the Lyapunov equation~\cite{Cernotik2015}
\begin{equation}
	\dot{\vect{V}} = \vect{AV}+\vect{VA}^T+\vect{N}.
\end{equation}
The covariance matrix has the elements $V_{ij} = \avg{r_ir_j+r_jr_i} - 2\avg{r_i}\avg{r_j}$ with $\vect{r} = (X,Y,x,p)^T$ and the cavity quadratures $X = (c+c^\dagger)/\sqrt{2}$, $Y = -i(a-a^\dagger)/\sqrt{2}$.
For the dynamics described by Eqs.~\eqref{eq:Langevin}, the drift and diffusion matrices are given by
\begin{subequations}
\begin{align}
	\vect{A} &= \left(\begin{array}{cccc}
		-\kappa & \Delta & 0 & 0 \\
		-\Delta & -\kappa & \sqrt{2}\lambda & 0\\
		0 & 0 & 0 &\omega_x \\
		\sqrt{2}\lambda & 0 & -\omega_x & -\gamma 
	\end{array}\right),\\
	\vect{N} &= \mathrm{diag}[2\kappa,2\kappa,0,2\gamma(2\nbar+1)];
\end{align}
\end{subequations}
generally, the drift matrix can be read off the Langevin equations and the diffusion matrix can be evaluated from the correlation functions of the input noises.

For steady-state generation of squeezing, dynamical stability of the system is crucial as a physical steady state does not exist in the unstable regime.
The stability can be assessed by evaluating the eigenvalues of the drift matrix $\vect{A}$;
for a stable system, all eigenvalues have negative real parts, signifying that none of the normal modes of the dynamics is amplified during evolution.

For the evaluation of mechanical squeezing, only the submatrix $\vect{V}_m$ describing the mechanical covariances is relevant,
which can be obtained from the block form
\begin{equation}
	\vect{V} = \left(\begin{array}{cc}
		\vect{V}_\mathrm{cav} & \vect{V}_C \\ \vect{V}_C^T & \vect{V}_m
	\end{array}\right)
\end{equation}
with $2\times 2$ blocks $\vect{V}_i$.
The amount of squeezing and antisqueezing is found by diagonalizing this matrix with the variances of the squeezed and antisqueezed quadratures given by
\begin{equation}
	V_\mathrm{sq} = \min\,\mathrm{eig}(\vect{V}_m),\qquad V_\mathrm{asq} = \max\,\mathrm{eig}(\vect{V}_m).
\end{equation}
The presence of squeezing is in general characterized by $\eta = V_\mathrm{sq}/V_\mathrm{asq} < 1$; nonclassical squeezing is present for $V_\mathrm{sq} < 1$.

\section{Amplitude modulation of the tweezer}\label{app:modulation}

\begin{figure}
	\centering
	\includegraphics[width=0.75\linewidth]{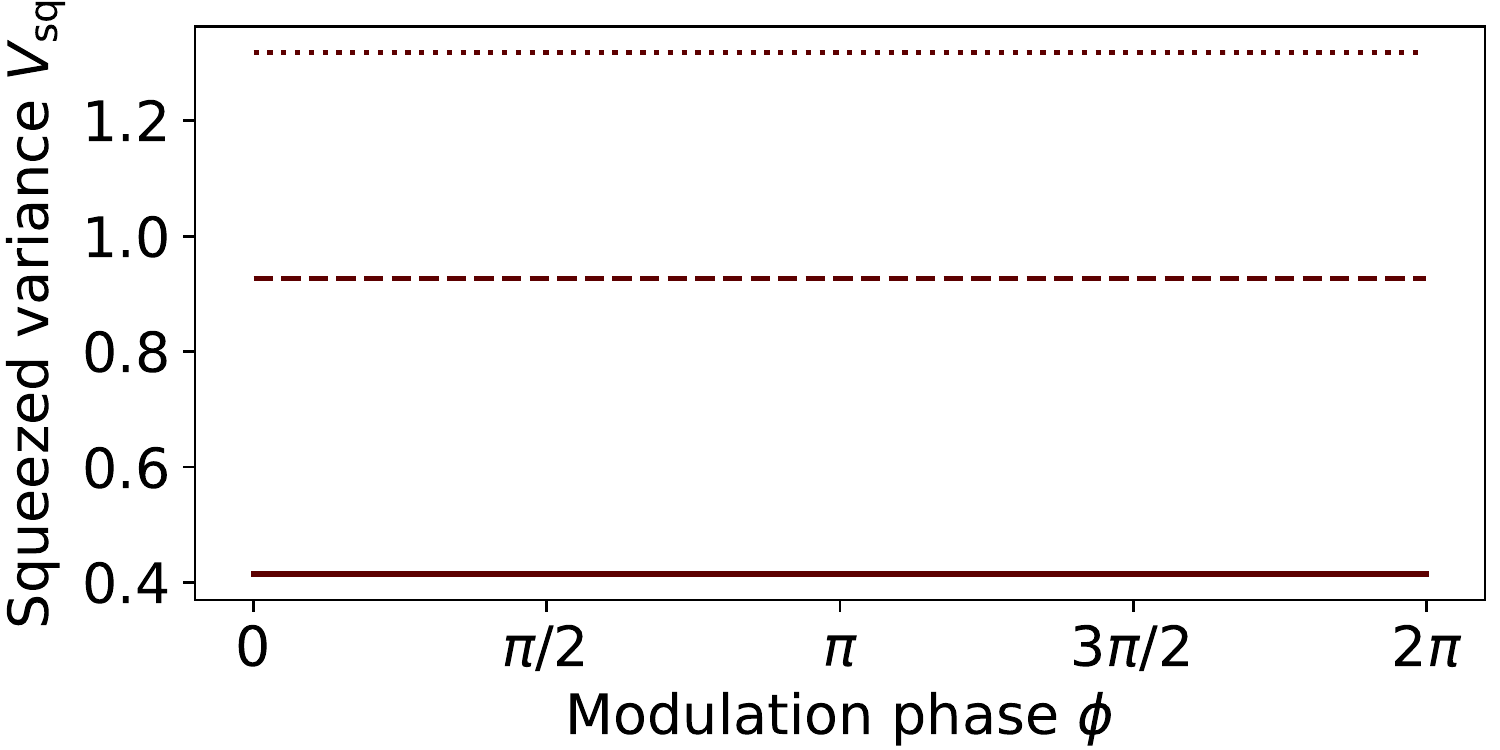}
	\caption{\label{fig:SMModPhase}
		Squeezed variance $\Vsq$ as a function of the modulation phase $\phi$.
		The modulation depths are $\alpha = 0.4$ (solid), $\alpha = 0.1$ (dashed), and $\alpha = 0.01$ (dotted);
		other parameters are the same as in Fig.~\ref{fig:Modulation} with $Q_m = \omega_x/\gamma_m = 10^9$.}
\end{figure}

To adiabatically eliminate the cavity dynamics in the case of modulated trapping beam, we move to the rotating frame with respect to $H_m = \frac{1}{2}\omega_x(p^2+x^2)$.
The cavity mode then obeys the equation of motion
\begin{widetext}
\begin{equation}
	\dot{c} = -(\kappa+i\Delta)c +\sqrt{2\kappa}c_\inpt +\frac{i\lambda}{\sqrt{2}}[1+\alpha\cos(2\omega_x t+\phi)](be^{-i\omega_xt}+b^\dagger e^{i\omega_xt}),
\end{equation}
where we introduced the mechanical annihilation operator $b = (x+ip)/\sqrt{2}$.
This equation can be integrated formally, assuming $b$ evolves slowly on the time scale of the cavity dynamics (given by $\Delta$).
We thus obtain
\begin{equation}
	c(t) = \sqrt{2\kappa}\int_0^td\tau\,e^{-(\kappa+i\Delta)(t-\tau)}c_\inpt(\tau)
	+\frac{i\lambda}{2}[1+\alpha\cos(2\omega_x t+\phi)]\left(\frac{be^{-i\omega_xt}}{\kappa+i(\Delta-\omega_x)}+\frac{b^\dagger e^{i\omega_xt}}{\kappa+i(\Delta+\omega_x)}\right).
\end{equation}
Plugging this result into the equations of motion for the particle, assuming $\Delta\gg\kappa\gg\omega_x$, and employing the rotating wave approximation, we get
\begin{subequations}\label{eq:SMLangEff}
\begin{align}
	\dot{x} &= \frac{\omega_x\Delta\alpha^2-2\lambda^2(2+\alpha^2)}{4\Delta}p -\frac{\omega_x\Delta\alpha-2\lambda^2\alpha}{2\Delta}(x\sin\phi+p\cos\phi)
		-\lambda\sqrt{2\kappa}\sin(\omega_xt)(\tilde{c}_\inpt+\tilde{c}_\inpt^\dagger),\\
	\dot{p} &= -\frac{\omega_x\Delta\alpha^2-2\lambda^2(2+\alpha^2)}{4\Delta}x-\frac{\omega_x\Delta\alpha-2\lambda^2\alpha}{2\Delta}(x\cos\phi-p\sin\phi)
		+\lambda\sqrt{2\kappa}\cos(\omega_xt)(\tilde{c}_\inpt+\tilde{c}_\inpt^\dagger).
\end{align}
\end{subequations}
Here, $\tilde{c}_\inpt = \int d\tau\,e^{-(\kappa+i\Delta)(t-\tau)}c_\inpt(t)$ is the optical input noise filtered by the cavity.

To bring the optical noise into the standard delta-correlated form, we employ the usual definition of the input field $c_\inpt(t) = (1/\sqrt{2\pi})\int d\omega\,e^{-i\omega t}c_0(\omega)$, where $c_0(\omega)$ is the initial state of the external mode $c(\omega)$ at time $t = 0$.
We can then write
\begin{equation}
	\tilde{c}_\inpt(t) = \frac{1}{\sqrt{2\pi}}\int d\omega\int d\tau\,e^{-(\kappa+i\Delta)(t-\tau)}e^{-i\omega\tau}c_0(\omega) 
	\approx -\frac{i}{\sqrt{2\pi}\Delta}\int d\omega\, e^{-i\omega t}c_0(\omega) 
	= -\frac{i}{\Delta}c_\inpt(t).
\end{equation}
Defining a rotating-frame noise operator $\bar{c}_\inpt = {c}_\inpt e^{i\omega_x t}$, we can now write (employing the rotating wave approximation)
\begin{subequations}
\begin{align}
	\sin(\omega_x t)(\tilde{c}_\inpt+\tilde{c}_\inpt^\dagger) &= -\frac{i}{\Delta}\frac{e^{i\omega_x t}-e^{-i\omega_x t}}{2i}(\bar{c}_\inpt e^{-i\omega_xt}-\bar{c}_\inpt^\dagger e^{i\omega_x t}) \approx -\frac{1}{2\Delta}(\bar{c}_\inpt+\bar{c}_\inpt^\dagger) = -\frac{1}{\sqrt{2}\Delta}\bar{X}_\inpt, \\
	\cos(\omega_x t)(\tilde{c}_\inpt + \tilde{c}_\inpt^\dagger) &\approx \frac{1}{\sqrt{2}\Delta}\bar{Y}_\inpt.
\end{align}
\end{subequations}
Eqs.~\eqref{eq:SMLangEff} can thus be simplified to
\begin{subequations}
\begin{align}
	\dot{x} &= \frac{\omega_x\Delta\alpha^2-2\lambda^2(2+\alpha^2)}{4\Delta}p -\frac{\omega_x\Delta\alpha-2\lambda^2\alpha}{2\Delta}(x\sin\phi+p\cos\phi) +\frac{\lambda\sqrt{\kappa}}{\Delta}\bar{X}_\inpt,\\
	\dot{p} &= -\frac{\omega_x\Delta\alpha^2-2\lambda^2(2+\alpha^2)}{4\Delta}x-\frac{\omega_x\Delta\alpha-2\lambda^2\alpha}{2\Delta}(x\cos\phi-p\sin\phi) +\frac{\lambda\sqrt{\kappa}}{\Delta}\bar{Y}_\inpt.
\end{align}
\end{subequations}
These equations are identical to Eqs.~\eqref{eq:LangEffective} with the noise operators $\bar{X}_\inpt,\bar{Y}_\inpt$ replaced by $X_\inpt,Y_\inpt$;
since their correlation properties are identical, this replacement does not affect our numerical analysis.

\section{Parametric and dissipative squeezing}\label{app:dissipative}

To squeeze the mechanical motion dissipatively, we start from the Hamiltonian for modulated tweezer, Eq.~\eqref{eq:SMModulated}, and move to the rotating frame with respect to the free Hamiltonian $H_0 = \frac{1}{2}\omega_x(p^2+x^2)+\Delta c^\dagger c$.
We thus obtain
\begin{align}
\begin{split}
    \Hint &= \frac{\omega_x\alpha}{4}(b^2 +b^{\dagger 2} + \alpha b^\dagger b) -\frac{\lambda}{\sqrt{2}}(b ce^{-i(\omega_x+\Delta)t} +bc^\dagger e^{-i(\omega_x-\Delta)t} +
        b^\dagger c^\dagger e^{i(\omega_x+\Delta)t} +b^\dagger c e^{i(\omega_x-\Delta)t}) \\
    &\quad -\frac{\lambda\alpha}{\sqrt{2}}\cos(2\omega_xt+\phi)[bce^{-i(\omega_x+\Delta)t}
        +bc^\dagger e^{-i(\omega_x-\Delta)t}+\Hc];
\end{split}
\end{align}
\end{widetext}
in deriving this expression, we applied the rotating frame to neglect terms oscillating with multiples of $\pm\omega_x t$.
With $\Delta = \omega_x$ and under the rotating wave approximation, the interaction Hamiltonian can be further simplified to
\begin{align}\label{eq:Squeezing1D}
	\Hint &= \frac{\omega_x\alpha}{4}(\alpha b^\dagger b+b^2e^{i\phi}-\frac{\lambda}{\sqrt{2}}\left(b+\frac{\alpha}{2}b^\dagger e^{-i\phi}\right)c^\dagger +\Hc\nonumber\\
	&= \frac{\omega_x\alpha}{4}(\beta^2-\alpha\beta^\dagger\beta)-\frac{\lambda}{2}\sqrt{\frac{4-\alpha^2}{2}}\beta c^\dagger +\Hc
\end{align}
In the second line, we introduced the mechanical Bogoliubov mode $\beta = (2b+\alpha b^\dagger)/\sqrt{4-\alpha^2}$ and took $\phi = 0$;
the modulation phase is irrelevant for dissipative squeezing as we show in Fig.~\ref{fig:SMModPhase}.

\begin{thebibliography}{62}%
\makeatletter
\providecommand \@ifxundefined [1]{%
 \@ifx{#1\undefined}
}%
\providecommand \@ifnum [1]{%
 \ifnum #1\expandafter \@firstoftwo
 \else \expandafter \@secondoftwo
 \fi
}%
\providecommand \@ifx [1]{%
 \ifx #1\expandafter \@firstoftwo
 \else \expandafter \@secondoftwo
 \fi
}%
\providecommand \natexlab [1]{#1}%
\providecommand \enquote  [1]{``#1''}%
\providecommand \bibnamefont  [1]{#1}%
\providecommand \bibfnamefont [1]{#1}%
\providecommand \citenamefont [1]{#1}%
\providecommand \href@noop [0]{\@secondoftwo}%
\providecommand \href [0]{\begingroup \@sanitize@url \@href}%
\providecommand \@href[1]{\@@startlink{#1}\@@href}%
\providecommand \@@href[1]{\endgroup#1\@@endlink}%
\providecommand \@sanitize@url [0]{\catcode `\\12\catcode `\$12\catcode
  `\&12\catcode `\#12\catcode `\^12\catcode `\_12\catcode `\%12\relax}%
\providecommand \@@startlink[1]{}%
\providecommand \@@endlink[0]{}%
\providecommand \url  [0]{\begingroup\@sanitize@url \@url }%
\providecommand \@url [1]{\endgroup\@href {#1}{\urlprefix }}%
\providecommand \urlprefix  [0]{URL }%
\providecommand \Eprint [0]{\href }%
\providecommand \doibase [0]{https://doi.org/}%
\providecommand \selectlanguage [0]{\@gobble}%
\providecommand \bibinfo  [0]{\@secondoftwo}%
\providecommand \bibfield  [0]{\@secondoftwo}%
\providecommand \translation [1]{[#1]}%
\providecommand \BibitemOpen [0]{}%
\providecommand \bibitemStop [0]{}%
\providecommand \bibitemNoStop [0]{.\EOS\space}%
\providecommand \EOS [0]{\spacefactor3000\relax}%
\providecommand \BibitemShut  [1]{\csname bibitem#1\endcsname}%
\let\auto@bib@innerbib\@empty
\bibitem [{\citenamefont {Aspelmeyer}\ \emph {et~al.}(2014)\citenamefont
  {Aspelmeyer}, \citenamefont {Kippenberg},\ and\ \citenamefont
  {Marquardt}}]{Aspelmeyer2014}%
  \BibitemOpen
  \bibfield  {author} {\bibinfo {author} {\bibfnamefont {M.}~\bibnamefont
  {Aspelmeyer}}, \bibinfo {author} {\bibfnamefont {T.~J.}\ \bibnamefont
  {Kippenberg}},\ and\ \bibinfo {author} {\bibfnamefont {F.}~\bibnamefont
  {Marquardt}},\ }\bibfield  {title} {\bibinfo {title} {{Cavity
  optomechanics}},\ }\href {https://doi.org/10.1103/RevModPhys.86.1391}
  {\bibfield  {journal} {\bibinfo  {journal} {Reviews of Modern Physics}\
  }\textbf {\bibinfo {volume} {86}},\ \bibinfo {pages} {1391} (\bibinfo {year}
  {2014})}\BibitemShut {NoStop}%
\bibitem [{\citenamefont {Forstner}\ \emph {et~al.}(2012)\citenamefont
  {Forstner}, \citenamefont {Prams}, \citenamefont {Knittel}, \citenamefont
  {van Ooijen}, \citenamefont {Swaim}, \citenamefont {Harris}, \citenamefont
  {Szorkovszky}, \citenamefont {Bowen},\ and\ \citenamefont
  {Rubinsztein-Dunlop}}]{Forstner2012}%
  \BibitemOpen
  \bibfield  {author} {\bibinfo {author} {\bibfnamefont {S.}~\bibnamefont
  {Forstner}}, \bibinfo {author} {\bibfnamefont {S.}~\bibnamefont {Prams}},
  \bibinfo {author} {\bibfnamefont {J.}~\bibnamefont {Knittel}}, \bibinfo
  {author} {\bibfnamefont {E.~D.}\ \bibnamefont {van Ooijen}}, \bibinfo
  {author} {\bibfnamefont {J.~D.}\ \bibnamefont {Swaim}}, \bibinfo {author}
  {\bibfnamefont {G.~I.}\ \bibnamefont {Harris}}, \bibinfo {author}
  {\bibfnamefont {A.}~\bibnamefont {Szorkovszky}}, \bibinfo {author}
  {\bibfnamefont {W.~P.}\ \bibnamefont {Bowen}},\ and\ \bibinfo {author}
  {\bibfnamefont {H.}~\bibnamefont {Rubinsztein-Dunlop}},\ }\bibfield  {title}
  {\bibinfo {title} {{Cavity Optomechanical Magnetometer}},\ }\href
  {https://doi.org/10.1103/PhysRevLett.108.120801} {\bibfield  {journal}
  {\bibinfo  {journal} {Physical Review Letters}\ }\textbf {\bibinfo {volume}
  {108}},\ \bibinfo {pages} {120801} (\bibinfo {year} {2012})}\BibitemShut
  {NoStop}%
\bibitem [{\citenamefont {Buchmann}\ \emph {et~al.}(2016)\citenamefont
  {Buchmann}, \citenamefont {Schreppler}, \citenamefont {Kohler}, \citenamefont
  {Spethmann},\ and\ \citenamefont {Stamper-Kurn}}]{Buchmann2016}%
  \BibitemOpen
  \bibfield  {author} {\bibinfo {author} {\bibfnamefont {L.~F.}\ \bibnamefont
  {Buchmann}}, \bibinfo {author} {\bibfnamefont {S.}~\bibnamefont
  {Schreppler}}, \bibinfo {author} {\bibfnamefont {J.}~\bibnamefont {Kohler}},
  \bibinfo {author} {\bibfnamefont {N.}~\bibnamefont {Spethmann}},\ and\
  \bibinfo {author} {\bibfnamefont {D.~M.}\ \bibnamefont {Stamper-Kurn}},\
  }\bibfield  {title} {\bibinfo {title} {{Complex Squeezing and Force
  Measurement Beyond the Standard Quantum Limit}},\ }\href
  {https://doi.org/10.1103/PhysRevLett.117.030801} {\bibfield  {journal}
  {\bibinfo  {journal} {Physical Review Letters}\ }\textbf {\bibinfo {volume}
  {117}},\ \bibinfo {pages} {030801} (\bibinfo {year} {2016})}\BibitemShut
  {NoStop}%
\bibitem [{\citenamefont {de~L{\'{e}}pinay}\ \emph {et~al.}(2017)\citenamefont
  {de~L{\'{e}}pinay}, \citenamefont {Pigeau}, \citenamefont {Besga},
  \citenamefont {Vincent}, \citenamefont {Poncharal},\ and\ \citenamefont
  {Arcizet}}]{DeLepinay2017}%
  \BibitemOpen
  \bibfield  {author} {\bibinfo {author} {\bibfnamefont {L.~M.}\ \bibnamefont
  {de~L{\'{e}}pinay}}, \bibinfo {author} {\bibfnamefont {B.}~\bibnamefont
  {Pigeau}}, \bibinfo {author} {\bibfnamefont {B.}~\bibnamefont {Besga}},
  \bibinfo {author} {\bibfnamefont {P.}~\bibnamefont {Vincent}}, \bibinfo
  {author} {\bibfnamefont {P.}~\bibnamefont {Poncharal}},\ and\ \bibinfo
  {author} {\bibfnamefont {O.}~\bibnamefont {Arcizet}},\ }\bibfield  {title}
  {\bibinfo {title} {{A universal and ultrasensitive vectorial nanomechanical
  sensor for imaging 2D force fields}},\ }\href
  {https://doi.org/10.1038/nnano.2016.193} {\bibfield  {journal} {\bibinfo
  {journal} {Nature Nanotechnology}\ }\textbf {\bibinfo {volume} {12}},\
  \bibinfo {pages} {156} (\bibinfo {year} {2017})}\BibitemShut {NoStop}%
\bibitem [{\citenamefont {Ockeloen-Korppi}\ \emph
  {et~al.}(2018{\natexlab{a}})\citenamefont {Ockeloen-Korppi}, \citenamefont
  {Damsk\"agg}, \citenamefont {Paraoanu}, \citenamefont {Massel},\ and\
  \citenamefont {Sillanp\"a\"a}}]{Ockeloen-Korppi2018a}%
  \BibitemOpen
  \bibfield  {author} {\bibinfo {author} {\bibfnamefont {C.~F.}\ \bibnamefont
  {Ockeloen-Korppi}}, \bibinfo {author} {\bibfnamefont {E.}~\bibnamefont
  {Damsk\"agg}}, \bibinfo {author} {\bibfnamefont {G.~S.}\ \bibnamefont
  {Paraoanu}}, \bibinfo {author} {\bibfnamefont {F.}~\bibnamefont {Massel}},\
  and\ \bibinfo {author} {\bibfnamefont {M.~A.}\ \bibnamefont
  {Sillanp\"a\"a}},\ }\bibfield  {title} {\bibinfo {title} {Revealing hidden
  quantum correlations in an electromechanical measurement},\ }\href
  {https://doi.org/10.1103/PhysRevLett.121.243601} {\bibfield  {journal}
  {\bibinfo  {journal} {Physical Review Letters}\ }\textbf {\bibinfo {volume}
  {121}},\ \bibinfo {pages} {243601} (\bibinfo {year}
  {2018}{\natexlab{a}})}\BibitemShut {NoStop}%
\bibitem [{\citenamefont {Pikovski}\ \emph {et~al.}(2012)\citenamefont
  {Pikovski}, \citenamefont {Vanner}, \citenamefont {Aspelmeyer}, \citenamefont
  {Kim},\ and\ \citenamefont {Brukner}}]{Pikovski2012}%
  \BibitemOpen
  \bibfield  {author} {\bibinfo {author} {\bibfnamefont {I.}~\bibnamefont
  {Pikovski}}, \bibinfo {author} {\bibfnamefont {M.~R.}\ \bibnamefont
  {Vanner}}, \bibinfo {author} {\bibfnamefont {M.}~\bibnamefont {Aspelmeyer}},
  \bibinfo {author} {\bibfnamefont {M.~S.}\ \bibnamefont {Kim}},\ and\ \bibinfo
  {author} {\bibfnamefont {{\v{C}}.}~\bibnamefont {Brukner}},\ }\bibfield
  {title} {\bibinfo {title} {{Probing Planck-scale physics with quantum
  optics}},\ }\href {https://doi.org/10.1038/nphys2262} {\bibfield  {journal}
  {\bibinfo  {journal} {Nature Physics}\ }\textbf {\bibinfo {volume} {8}},\
  \bibinfo {pages} {393} (\bibinfo {year} {2012})}\BibitemShut {NoStop}%
\bibitem [{\citenamefont {Pfister}\ \emph {et~al.}(2016)\citenamefont
  {Pfister}, \citenamefont {Kaniewski}, \citenamefont {Tomamichel},
  \citenamefont {Mantri}, \citenamefont {Schmucker}, \citenamefont {McMahon},
  \citenamefont {Milburn},\ and\ \citenamefont {Wehner}}]{Pfister2016}%
  \BibitemOpen
  \bibfield  {author} {\bibinfo {author} {\bibfnamefont {C.}~\bibnamefont
  {Pfister}}, \bibinfo {author} {\bibfnamefont {J.}~\bibnamefont {Kaniewski}},
  \bibinfo {author} {\bibfnamefont {M.}~\bibnamefont {Tomamichel}}, \bibinfo
  {author} {\bibfnamefont {A.}~\bibnamefont {Mantri}}, \bibinfo {author}
  {\bibfnamefont {R.}~\bibnamefont {Schmucker}}, \bibinfo {author}
  {\bibfnamefont {N.}~\bibnamefont {McMahon}}, \bibinfo {author} {\bibfnamefont
  {G.}~\bibnamefont {Milburn}},\ and\ \bibinfo {author} {\bibfnamefont
  {S.}~\bibnamefont {Wehner}},\ }\bibfield  {title} {\bibinfo {title} {{A
  universal test for gravitational decoherence}},\ }\href
  {https://doi.org/10.1038/ncomms13022} {\bibfield  {journal} {\bibinfo
  {journal} {Nature Communications}\ }\textbf {\bibinfo {volume} {7}},\
  \bibinfo {pages} {13022} (\bibinfo {year} {2016})}\BibitemShut {NoStop}%
\bibitem [{\citenamefont {Chang}\ \emph {et~al.}(2009)\citenamefont {Chang},
  \citenamefont {Regal}, \citenamefont {Papp}, \citenamefont {Wilson},
  \citenamefont {Ye}, \citenamefont {Painter}, \citenamefont {Kimble},\ and\
  \citenamefont {Zoller}}]{Chang2009}%
  \BibitemOpen
  \bibfield  {author} {\bibinfo {author} {\bibfnamefont {D.~E.}\ \bibnamefont
  {Chang}}, \bibinfo {author} {\bibfnamefont {C.~A.}\ \bibnamefont {Regal}},
  \bibinfo {author} {\bibfnamefont {S.~B.}\ \bibnamefont {Papp}}, \bibinfo
  {author} {\bibfnamefont {D.~J.}\ \bibnamefont {Wilson}}, \bibinfo {author}
  {\bibfnamefont {J.}~\bibnamefont {Ye}}, \bibinfo {author} {\bibfnamefont
  {O.}~\bibnamefont {Painter}}, \bibinfo {author} {\bibfnamefont {H.~J.}\
  \bibnamefont {Kimble}},\ and\ \bibinfo {author} {\bibfnamefont
  {P.}~\bibnamefont {Zoller}},\ }\bibfield  {title} {\bibinfo {title} {Cavity
  opto-mechanics using an optically levitated nanosphere},\ }\href
  {https://doi.org/10.1073/pnas.0912969107} {\bibfield  {journal} {\bibinfo
  {journal} {Proceedings of the National Academy of Sciences}\ }\textbf
  {\bibinfo {volume} {107}},\ \bibinfo {pages} {1005} (\bibinfo {year}
  {2009})}\BibitemShut {NoStop}%
\bibitem [{\citenamefont {Romero-Isart}\ \emph
  {et~al.}(2011{\natexlab{a}})\citenamefont {Romero-Isart}, \citenamefont
  {Pflanzer}, \citenamefont {Juan}, \citenamefont {Quidant}, \citenamefont
  {Kiesel}, \citenamefont {Aspelmeyer},\ and\ \citenamefont
  {Cirac}}]{Romero-Isart2011}%
  \BibitemOpen
  \bibfield  {author} {\bibinfo {author} {\bibfnamefont {O.}~\bibnamefont
  {Romero-Isart}}, \bibinfo {author} {\bibfnamefont {A.~C.}\ \bibnamefont
  {Pflanzer}}, \bibinfo {author} {\bibfnamefont {M.~L.}\ \bibnamefont {Juan}},
  \bibinfo {author} {\bibfnamefont {R.}~\bibnamefont {Quidant}}, \bibinfo
  {author} {\bibfnamefont {N.}~\bibnamefont {Kiesel}}, \bibinfo {author}
  {\bibfnamefont {M.}~\bibnamefont {Aspelmeyer}},\ and\ \bibinfo {author}
  {\bibfnamefont {J.~I.}\ \bibnamefont {Cirac}},\ }\bibfield  {title} {\bibinfo
  {title} {Optically levitating dielectrics in the quantum regime: Theory and
  protocols},\ }\href {https://doi.org/10.1103/PhysRevA.83.013803} {\bibfield
  {journal} {\bibinfo  {journal} {Physical Review A}\ }\textbf {\bibinfo
  {volume} {83}},\ \bibinfo {pages} {013803} (\bibinfo {year}
  {2011}{\natexlab{a}})}\BibitemShut {NoStop}%
\bibitem [{\citenamefont {Millen}\ \emph {et~al.}(2019)\citenamefont {Millen},
  \citenamefont {Monteiro}, \citenamefont {Pettit},\ and\ \citenamefont
  {Vamivakas}}]{Millen2019}%
  \BibitemOpen
  \bibfield  {author} {\bibinfo {author} {\bibfnamefont {J.}~\bibnamefont
  {Millen}}, \bibinfo {author} {\bibfnamefont {T.~S.}\ \bibnamefont
  {Monteiro}}, \bibinfo {author} {\bibfnamefont {R.}~\bibnamefont {Pettit}},\
  and\ \bibinfo {author} {\bibfnamefont {A.~N.}\ \bibnamefont {Vamivakas}},\
  }\bibfield  {title} {\bibinfo {title} {Optomechanics with levitated
  particles},\ }\href {https://arxiv.org/abs/1907.08198} {\  (\bibinfo {year}
  {2019})},\ \Eprint {https://arxiv.org/abs/arXiv:1907.08198}
  {arXiv:1907.08198} \BibitemShut {NoStop}%
\bibitem [{\citenamefont {Prat-Camps}\ \emph {et~al.}(2017)\citenamefont
  {Prat-Camps}, \citenamefont {Teo}, \citenamefont {Rusconi}, \citenamefont
  {Wieczorek},\ and\ \citenamefont {Romero-Isart}}]{Prat-Camps2017}%
  \BibitemOpen
  \bibfield  {author} {\bibinfo {author} {\bibfnamefont {J.}~\bibnamefont
  {Prat-Camps}}, \bibinfo {author} {\bibfnamefont {C.}~\bibnamefont {Teo}},
  \bibinfo {author} {\bibfnamefont {C.~C.}\ \bibnamefont {Rusconi}}, \bibinfo
  {author} {\bibfnamefont {W.}~\bibnamefont {Wieczorek}},\ and\ \bibinfo
  {author} {\bibfnamefont {O.}~\bibnamefont {Romero-Isart}},\ }\bibfield
  {title} {\bibinfo {title} {{Ultrasensitive Inertial and Force Sensors with
  Diamagnetically Levitated Magnets}},\ }\href
  {https://doi.org/10.1103/PhysRevApplied.8.034002} {\bibfield  {journal}
  {\bibinfo  {journal} {Physical Review Applied}\ }\textbf {\bibinfo {volume}
  {8}},\ \bibinfo {pages} {034002} (\bibinfo {year} {2017})}\BibitemShut
  {NoStop}%
\bibitem [{\citenamefont {Hebestreit}\ \emph {et~al.}(2018)\citenamefont
  {Hebestreit}, \citenamefont {Frimmer}, \citenamefont {Reimann},\ and\
  \citenamefont {Novotny}}]{Hebestreit2018}%
  \BibitemOpen
  \bibfield  {author} {\bibinfo {author} {\bibfnamefont {E.}~\bibnamefont
  {Hebestreit}}, \bibinfo {author} {\bibfnamefont {M.}~\bibnamefont {Frimmer}},
  \bibinfo {author} {\bibfnamefont {R.}~\bibnamefont {Reimann}},\ and\ \bibinfo
  {author} {\bibfnamefont {L.}~\bibnamefont {Novotny}},\ }\bibfield  {title}
  {\bibinfo {title} {{Sensing Static Forces with Free-Falling Nanoparticles}},\
  }\href {https://doi.org/10.1103/PhysRevLett.121.063602} {\bibfield  {journal}
  {\bibinfo  {journal} {Physical Review Letters}\ }\textbf {\bibinfo {volume}
  {121}},\ \bibinfo {pages} {063602} (\bibinfo {year} {2018})}\BibitemShut
  {NoStop}%
\bibitem [{\citenamefont {Blakemore}\ \emph {et~al.}(2019)\citenamefont
  {Blakemore}, \citenamefont {Rider}, \citenamefont {Roy}, \citenamefont
  {Wang}, \citenamefont {Kawasaki},\ and\ \citenamefont
  {Gratta}}]{Blakemore2019}%
  \BibitemOpen
  \bibfield  {author} {\bibinfo {author} {\bibfnamefont {C.~P.}\ \bibnamefont
  {Blakemore}}, \bibinfo {author} {\bibfnamefont {A.~D.}\ \bibnamefont
  {Rider}}, \bibinfo {author} {\bibfnamefont {S.}~\bibnamefont {Roy}}, \bibinfo
  {author} {\bibfnamefont {Q.}~\bibnamefont {Wang}}, \bibinfo {author}
  {\bibfnamefont {A.}~\bibnamefont {Kawasaki}},\ and\ \bibinfo {author}
  {\bibfnamefont {G.}~\bibnamefont {Gratta}},\ }\bibfield  {title} {\bibinfo
  {title} {Three dimensional force-field microscopy with optically levitated
  microspheres},\ }\href {https://doi.org/10.1103/PhysRevA.99.023816}
  {\bibfield  {journal} {\bibinfo  {journal} {Physical Review A}\ }\textbf
  {\bibinfo {volume} {99}},\ \bibinfo {pages} {023816} (\bibinfo {year}
  {2019})}\BibitemShut {NoStop}%
\bibitem [{\citenamefont {Gieseler}\ and\ \citenamefont
  {Millen}(2018)}]{Gieseler2018}%
  \BibitemOpen
  \bibfield  {author} {\bibinfo {author} {\bibfnamefont {J.}~\bibnamefont
  {Gieseler}}\ and\ \bibinfo {author} {\bibfnamefont {J.}~\bibnamefont
  {Millen}},\ }\bibfield  {title} {\bibinfo {title} {Levitated nanoparticles
  for microscopic thermodynamics---a review},\ }\href
  {https://doi.org/10.3390/e20050326} {\bibfield  {journal} {\bibinfo
  {journal} {Entropy}\ }\textbf {\bibinfo {volume} {20}},\ \bibinfo {pages}
  {326} (\bibinfo {year} {2018})}\BibitemShut {NoStop}%
\bibitem [{\citenamefont {{\v{S}}iler}\ \emph {et~al.}(2018)\citenamefont
  {{\v{S}}iler}, \citenamefont {Ornigotti}, \citenamefont {Brzobohat\'y},
  \citenamefont {J\'akl}, \citenamefont {Ryabov}, \citenamefont {Holubec},
  \citenamefont {Zem\'anek},\ and\ \citenamefont {Filip}}]{Siler2018}%
  \BibitemOpen
  \bibfield  {author} {\bibinfo {author} {\bibfnamefont {M.}~\bibnamefont
  {{\v{S}}iler}}, \bibinfo {author} {\bibfnamefont {L.}~\bibnamefont
  {Ornigotti}}, \bibinfo {author} {\bibfnamefont {O.}~\bibnamefont
  {Brzobohat\'y}}, \bibinfo {author} {\bibfnamefont {P.}~\bibnamefont
  {J\'akl}}, \bibinfo {author} {\bibfnamefont {A.}~\bibnamefont {Ryabov}},
  \bibinfo {author} {\bibfnamefont {V.}~\bibnamefont {Holubec}}, \bibinfo
  {author} {\bibfnamefont {P.}~\bibnamefont {Zem\'anek}},\ and\ \bibinfo
  {author} {\bibfnamefont {R.}~\bibnamefont {Filip}},\ }\bibfield  {title}
  {\bibinfo {title} {{Diffusing up the Hill: Dynamics and Equipartition in
  Highly Unstable Systems}},\ }\href
  {https://doi.org/10.1103/PhysRevLett.121.230601} {\bibfield  {journal}
  {\bibinfo  {journal} {Physical Review Letters}\ }\textbf {\bibinfo {volume}
  {121}},\ \bibinfo {pages} {230601} (\bibinfo {year} {2018})}\BibitemShut
  {NoStop}%
\bibitem [{\citenamefont {Romero-Isart}\ \emph
  {et~al.}(2011{\natexlab{b}})\citenamefont {Romero-Isart}, \citenamefont
  {Pflanzer}, \citenamefont {Blaser}, \citenamefont {Kaltenbaek}, \citenamefont
  {Kiesel}, \citenamefont {Aspelmeyer},\ and\ \citenamefont
  {Cirac}}]{Romero-Isart2011a}%
  \BibitemOpen
  \bibfield  {author} {\bibinfo {author} {\bibfnamefont {O.}~\bibnamefont
  {Romero-Isart}}, \bibinfo {author} {\bibfnamefont {A.~C.}\ \bibnamefont
  {Pflanzer}}, \bibinfo {author} {\bibfnamefont {F.}~\bibnamefont {Blaser}},
  \bibinfo {author} {\bibfnamefont {R.}~\bibnamefont {Kaltenbaek}}, \bibinfo
  {author} {\bibfnamefont {N.}~\bibnamefont {Kiesel}}, \bibinfo {author}
  {\bibfnamefont {M.}~\bibnamefont {Aspelmeyer}},\ and\ \bibinfo {author}
  {\bibfnamefont {J.~I.}\ \bibnamefont {Cirac}},\ }\bibfield  {title} {\bibinfo
  {title} {{Large Quantum Superpositions and Interference of Massive
  Nanometer-Sized Objects}},\ }\href
  {https://doi.org/10.1103/PhysRevLett.107.020405} {\bibfield  {journal}
  {\bibinfo  {journal} {Physical Review Letters}\ }\textbf {\bibinfo {volume}
  {107}},\ \bibinfo {pages} {020405} (\bibinfo {year}
  {2011}{\natexlab{b}})}\BibitemShut {NoStop}%
\bibitem [{\citenamefont {Bateman}\ \emph {et~al.}(2014)\citenamefont
  {Bateman}, \citenamefont {Nimmrichter}, \citenamefont {Hornberger},\ and\
  \citenamefont {Ulbricht}}]{Bateman2014}%
  \BibitemOpen
  \bibfield  {author} {\bibinfo {author} {\bibfnamefont {J.}~\bibnamefont
  {Bateman}}, \bibinfo {author} {\bibfnamefont {S.}~\bibnamefont
  {Nimmrichter}}, \bibinfo {author} {\bibfnamefont {K.}~\bibnamefont
  {Hornberger}},\ and\ \bibinfo {author} {\bibfnamefont {H.}~\bibnamefont
  {Ulbricht}},\ }\bibfield  {title} {\bibinfo {title} {Near-field
  interferometry of a free-falling nanoparticle from a point-like source},\
  }\href {https://doi.org/10.1038/ncomms5788} {\bibfield  {journal} {\bibinfo
  {journal} {Nature Communications}\ }\textbf {\bibinfo {volume} {5}},\
  \bibinfo {pages} {4788} (\bibinfo {year} {2014})}\BibitemShut {NoStop}%
\bibitem [{\citenamefont {Moore}\ \emph {et~al.}(2014)\citenamefont {Moore},
  \citenamefont {Rider},\ and\ \citenamefont {Gratta}}]{Moore2014}%
  \BibitemOpen
  \bibfield  {author} {\bibinfo {author} {\bibfnamefont {D.~C.}\ \bibnamefont
  {Moore}}, \bibinfo {author} {\bibfnamefont {A.~D.}\ \bibnamefont {Rider}},\
  and\ \bibinfo {author} {\bibfnamefont {G.}~\bibnamefont {Gratta}},\
  }\bibfield  {title} {\bibinfo {title} {{Search for Millicharged Particles
  Using Optically Levitated Microspheres}},\ }\href
  {https://doi.org/10.1103/PhysRevLett.113.251801} {\bibfield  {journal}
  {\bibinfo  {journal} {Physical Review Letters}\ }\textbf {\bibinfo {volume}
  {113}},\ \bibinfo {pages} {251801} (\bibinfo {year} {2014})}\BibitemShut
  {NoStop}%
\bibitem [{\citenamefont {Rider}\ \emph {et~al.}(2016)\citenamefont {Rider},
  \citenamefont {Moore}, \citenamefont {Blakemore}, \citenamefont {Louis},
  \citenamefont {Lu},\ and\ \citenamefont {Gratta}}]{Rider2016}%
  \BibitemOpen
  \bibfield  {author} {\bibinfo {author} {\bibfnamefont {A.~D.}\ \bibnamefont
  {Rider}}, \bibinfo {author} {\bibfnamefont {D.~C.}\ \bibnamefont {Moore}},
  \bibinfo {author} {\bibfnamefont {C.~P.}\ \bibnamefont {Blakemore}}, \bibinfo
  {author} {\bibfnamefont {M.}~\bibnamefont {Louis}}, \bibinfo {author}
  {\bibfnamefont {M.}~\bibnamefont {Lu}},\ and\ \bibinfo {author}
  {\bibfnamefont {G.}~\bibnamefont {Gratta}},\ }\bibfield  {title} {\bibinfo
  {title} {{Search for Screened Interactions Associated with Dark Energy below
  the $100\text{ }\ensuremath{\mu}\mathrm{m}$ Length Scale}},\ }\href
  {https://doi.org/10.1103/PhysRevLett.117.101101} {\bibfield  {journal}
  {\bibinfo  {journal} {Physical Review Letters}\ }\textbf {\bibinfo {volume}
  {117}},\ \bibinfo {pages} {101101} (\bibinfo {year} {2016})}\BibitemShut
  {NoStop}%
\bibitem [{\citenamefont {Gieseler}\ \emph {et~al.}(2012)\citenamefont
  {Gieseler}, \citenamefont {Deutsch}, \citenamefont {Quidant},\ and\
  \citenamefont {Novotny}}]{Gieseler2012}%
  \BibitemOpen
  \bibfield  {author} {\bibinfo {author} {\bibfnamefont {J.}~\bibnamefont
  {Gieseler}}, \bibinfo {author} {\bibfnamefont {B.}~\bibnamefont {Deutsch}},
  \bibinfo {author} {\bibfnamefont {R.}~\bibnamefont {Quidant}},\ and\ \bibinfo
  {author} {\bibfnamefont {L.}~\bibnamefont {Novotny}},\ }\bibfield  {title}
  {\bibinfo {title} {{Subkelvin Parametric Feedback Cooling of a Laser-Trapped
  Nanoparticle}},\ }\href {https://doi.org/10.1103/PhysRevLett.109.103603}
  {\bibfield  {journal} {\bibinfo  {journal} {Physical Review Letters}\
  }\textbf {\bibinfo {volume} {109}},\ \bibinfo {pages} {103603} (\bibinfo
  {year} {2012})}\BibitemShut {NoStop}%
\bibitem [{\citenamefont {Kiesel}\ \emph {et~al.}(2013)\citenamefont {Kiesel},
  \citenamefont {Blaser}, \citenamefont {Delic}, \citenamefont {Grass},
  \citenamefont {Kaltenbaek},\ and\ \citenamefont {Aspelmeyer}}]{Kiesel2013}%
  \BibitemOpen
  \bibfield  {author} {\bibinfo {author} {\bibfnamefont {N.}~\bibnamefont
  {Kiesel}}, \bibinfo {author} {\bibfnamefont {F.}~\bibnamefont {Blaser}},
  \bibinfo {author} {\bibfnamefont {U.}~\bibnamefont {Delic}}, \bibinfo
  {author} {\bibfnamefont {D.}~\bibnamefont {Grass}}, \bibinfo {author}
  {\bibfnamefont {R.}~\bibnamefont {Kaltenbaek}},\ and\ \bibinfo {author}
  {\bibfnamefont {M.}~\bibnamefont {Aspelmeyer}},\ }\bibfield  {title}
  {\bibinfo {title} {Cavity cooling of an optically levitated submicron
  particle},\ }\href {https://doi.org/10.1073/pnas.1309167110} {\bibfield
  {journal} {\bibinfo  {journal} {Proceedings of the National Academy of
  Sciences}\ }\textbf {\bibinfo {volume} {110}},\ \bibinfo {pages} {14180}
  (\bibinfo {year} {2013})}\BibitemShut {NoStop}%
\bibitem [{\citenamefont {Asenbaum}\ \emph {et~al.}(2013)\citenamefont
  {Asenbaum}, \citenamefont {Kuhn}, \citenamefont {Nimmrichter}, \citenamefont
  {Sezer},\ and\ \citenamefont {Arndt}}]{Asenbaum2013}%
  \BibitemOpen
  \bibfield  {author} {\bibinfo {author} {\bibfnamefont {P.}~\bibnamefont
  {Asenbaum}}, \bibinfo {author} {\bibfnamefont {S.}~\bibnamefont {Kuhn}},
  \bibinfo {author} {\bibfnamefont {S.}~\bibnamefont {Nimmrichter}}, \bibinfo
  {author} {\bibfnamefont {U.}~\bibnamefont {Sezer}},\ and\ \bibinfo {author}
  {\bibfnamefont {M.}~\bibnamefont {Arndt}},\ }\bibfield  {title} {\bibinfo
  {title} {Cavity cooling of free silicon nanoparticles in high vacuum},\
  }\href {https://doi.org/10.1038/ncomms3743} {\bibfield  {journal} {\bibinfo
  {journal} {Nature Communications}\ }\textbf {\bibinfo {volume} {4}},\
  \bibinfo {pages} {2743} (\bibinfo {year} {2013})}\BibitemShut {NoStop}%
\bibitem [{\citenamefont {Conangla}\ \emph {et~al.}(2019)\citenamefont
  {Conangla}, \citenamefont {Ricci}, \citenamefont {Cuairan}, \citenamefont
  {Schell}, \citenamefont {Meyer},\ and\ \citenamefont
  {Quidant}}]{Conangla2019}%
  \BibitemOpen
  \bibfield  {author} {\bibinfo {author} {\bibfnamefont {G.~P.}\ \bibnamefont
  {Conangla}}, \bibinfo {author} {\bibfnamefont {F.}~\bibnamefont {Ricci}},
  \bibinfo {author} {\bibfnamefont {M.~T.}\ \bibnamefont {Cuairan}}, \bibinfo
  {author} {\bibfnamefont {A.~W.}\ \bibnamefont {Schell}}, \bibinfo {author}
  {\bibfnamefont {N.}~\bibnamefont {Meyer}},\ and\ \bibinfo {author}
  {\bibfnamefont {R.}~\bibnamefont {Quidant}},\ }\bibfield  {title} {\bibinfo
  {title} {{Optimal Feedback Cooling of a Charged Levitated Nanoparticle with
  Adaptive Control}},\ }\href {https://doi.org/10.1103/physrevlett.122.223602}
  {\bibfield  {journal} {\bibinfo  {journal} {Physical Review Letters}\
  }\textbf {\bibinfo {volume} {122}},\ \bibinfo {pages} {223602} (\bibinfo
  {year} {2019})}\BibitemShut {NoStop}%
\bibitem [{\citenamefont {Tebbenjohanns}\ \emph {et~al.}(2019)\citenamefont
  {Tebbenjohanns}, \citenamefont {Frimmer}, \citenamefont {Militaru},
  \citenamefont {Jain},\ and\ \citenamefont {Novotny}}]{Tebbenjohanns2019}%
  \BibitemOpen
  \bibfield  {author} {\bibinfo {author} {\bibfnamefont {F.}~\bibnamefont
  {Tebbenjohanns}}, \bibinfo {author} {\bibfnamefont {M.}~\bibnamefont
  {Frimmer}}, \bibinfo {author} {\bibfnamefont {A.}~\bibnamefont {Militaru}},
  \bibinfo {author} {\bibfnamefont {V.}~\bibnamefont {Jain}},\ and\ \bibinfo
  {author} {\bibfnamefont {L.}~\bibnamefont {Novotny}},\ }\bibfield  {title}
  {\bibinfo {title} {{Cold Damping of an Optically Levitated Nanoparticle to
  Microkelvin Temperatures}},\ }\href
  {https://doi.org/10.1103/physrevlett.122.223601} {\bibfield  {journal}
  {\bibinfo  {journal} {Physical Review Letters}\ }\textbf {\bibinfo {volume}
  {122}},\ \bibinfo {pages} {223601} (\bibinfo {year} {2019})}\BibitemShut
  {NoStop}%
\bibitem [{\citenamefont {Rashid}\ \emph {et~al.}(2016)\citenamefont {Rashid},
  \citenamefont {Tufarelli}, \citenamefont {Bateman}, \citenamefont {Vovrosh},
  \citenamefont {Hempston}, \citenamefont {Kim},\ and\ \citenamefont
  {Ulbricht}}]{Rashid2016}%
  \BibitemOpen
  \bibfield  {author} {\bibinfo {author} {\bibfnamefont {M.}~\bibnamefont
  {Rashid}}, \bibinfo {author} {\bibfnamefont {T.}~\bibnamefont {Tufarelli}},
  \bibinfo {author} {\bibfnamefont {J.}~\bibnamefont {Bateman}}, \bibinfo
  {author} {\bibfnamefont {J.}~\bibnamefont {Vovrosh}}, \bibinfo {author}
  {\bibfnamefont {D.}~\bibnamefont {Hempston}}, \bibinfo {author}
  {\bibfnamefont {M.~S.}\ \bibnamefont {Kim}},\ and\ \bibinfo {author}
  {\bibfnamefont {H.}~\bibnamefont {Ulbricht}},\ }\bibfield  {title} {\bibinfo
  {title} {{Experimental Realization of a Thermal Squeezed State of Levitated
  Optomechanics}},\ }\href {https://doi.org/10.1103/PhysRevLett.117.273601}
  {\bibfield  {journal} {\bibinfo  {journal} {Physical Review Letters}\
  }\textbf {\bibinfo {volume} {117}},\ \bibinfo {pages} {273601} (\bibinfo
  {year} {2016})}\BibitemShut {NoStop}%
\bibitem [{\citenamefont {Ahn}\ \emph {et~al.}(2018)\citenamefont {Ahn},
  \citenamefont {Xu}, \citenamefont {Bang}, \citenamefont {Deng}, \citenamefont
  {Hoang}, \citenamefont {Han}, \citenamefont {Ma},\ and\ \citenamefont
  {Li}}]{Ahn2018}%
  \BibitemOpen
  \bibfield  {author} {\bibinfo {author} {\bibfnamefont {J.}~\bibnamefont
  {Ahn}}, \bibinfo {author} {\bibfnamefont {Z.}~\bibnamefont {Xu}}, \bibinfo
  {author} {\bibfnamefont {J.}~\bibnamefont {Bang}}, \bibinfo {author}
  {\bibfnamefont {Y.-H.}\ \bibnamefont {Deng}}, \bibinfo {author}
  {\bibfnamefont {T.~M.}\ \bibnamefont {Hoang}}, \bibinfo {author}
  {\bibfnamefont {Q.}~\bibnamefont {Han}}, \bibinfo {author} {\bibfnamefont
  {R.-M.}\ \bibnamefont {Ma}},\ and\ \bibinfo {author} {\bibfnamefont
  {T.}~\bibnamefont {Li}},\ }\bibfield  {title} {\bibinfo {title} {{Optically
  Levitated Nanodumbbell Torsion Balance and GHz Nanomechanical Rotor}},\
  }\href {https://doi.org/10.1103/PhysRevLett.121.033603} {\bibfield  {journal}
  {\bibinfo  {journal} {Physical Review Letters}\ }\textbf {\bibinfo {volume}
  {121}},\ \bibinfo {pages} {033603} (\bibinfo {year} {2018})}\BibitemShut
  {NoStop}%
\bibitem [{\citenamefont {Reimann}\ \emph {et~al.}(2018)\citenamefont
  {Reimann}, \citenamefont {Doderer}, \citenamefont {Hebestreit}, \citenamefont
  {Diehl}, \citenamefont {Frimmer}, \citenamefont {Windey}, \citenamefont
  {Tebbenjohanns},\ and\ \citenamefont {Novotny}}]{Reimann2018}%
  \BibitemOpen
  \bibfield  {author} {\bibinfo {author} {\bibfnamefont {R.}~\bibnamefont
  {Reimann}}, \bibinfo {author} {\bibfnamefont {M.}~\bibnamefont {Doderer}},
  \bibinfo {author} {\bibfnamefont {E.}~\bibnamefont {Hebestreit}}, \bibinfo
  {author} {\bibfnamefont {R.}~\bibnamefont {Diehl}}, \bibinfo {author}
  {\bibfnamefont {M.}~\bibnamefont {Frimmer}}, \bibinfo {author} {\bibfnamefont
  {D.}~\bibnamefont {Windey}}, \bibinfo {author} {\bibfnamefont
  {F.}~\bibnamefont {Tebbenjohanns}},\ and\ \bibinfo {author} {\bibfnamefont
  {L.}~\bibnamefont {Novotny}},\ }\bibfield  {title} {\bibinfo {title} {{GHz
  Rotation of an Optically Trapped Nanoparticle in Vacuum}},\ }\href
  {https://doi.org/10.1103/PhysRevLett.121.033602} {\bibfield  {journal}
  {\bibinfo  {journal} {Physical Review Letters}\ }\textbf {\bibinfo {volume}
  {121}},\ \bibinfo {pages} {033602} (\bibinfo {year} {2018})}\BibitemShut
  {NoStop}%
\bibitem [{\citenamefont {Hoang}\ \emph {et~al.}(2016)\citenamefont {Hoang},
  \citenamefont {Ma}, \citenamefont {Ahn}, \citenamefont {Bang}, \citenamefont
  {Robicheaux}, \citenamefont {Yin},\ and\ \citenamefont {Li}}]{Hoang2016}%
  \BibitemOpen
  \bibfield  {author} {\bibinfo {author} {\bibfnamefont {T.~M.}\ \bibnamefont
  {Hoang}}, \bibinfo {author} {\bibfnamefont {Y.}~\bibnamefont {Ma}}, \bibinfo
  {author} {\bibfnamefont {J.}~\bibnamefont {Ahn}}, \bibinfo {author}
  {\bibfnamefont {J.}~\bibnamefont {Bang}}, \bibinfo {author} {\bibfnamefont
  {F.}~\bibnamefont {Robicheaux}}, \bibinfo {author} {\bibfnamefont {Z.-Q.}\
  \bibnamefont {Yin}},\ and\ \bibinfo {author} {\bibfnamefont {T.}~\bibnamefont
  {Li}},\ }\bibfield  {title} {\bibinfo {title} {{Torsional Optomechanics of a
  Levitated Nonspherical Nanoparticle}},\ }\href
  {https://doi.org/10.1103/PhysRevLett.117.123604} {\bibfield  {journal}
  {\bibinfo  {journal} {Physical Review Letters}\ }\textbf {\bibinfo {volume}
  {117}},\ \bibinfo {pages} {123604} (\bibinfo {year} {2016})}\BibitemShut
  {NoStop}%
\bibitem [{\citenamefont {Kuhn}\ \emph {et~al.}(2017)\citenamefont {Kuhn},
  \citenamefont {Kosloff}, \citenamefont {Stickler}, \citenamefont {Patolsky},
  \citenamefont {Hornberger}, \citenamefont {Arndt},\ and\ \citenamefont
  {Millen}}]{Kuhn2017}%
  \BibitemOpen
  \bibfield  {author} {\bibinfo {author} {\bibfnamefont {S.}~\bibnamefont
  {Kuhn}}, \bibinfo {author} {\bibfnamefont {A.}~\bibnamefont {Kosloff}},
  \bibinfo {author} {\bibfnamefont {B.~A.}\ \bibnamefont {Stickler}}, \bibinfo
  {author} {\bibfnamefont {F.}~\bibnamefont {Patolsky}}, \bibinfo {author}
  {\bibfnamefont {K.}~\bibnamefont {Hornberger}}, \bibinfo {author}
  {\bibfnamefont {M.}~\bibnamefont {Arndt}},\ and\ \bibinfo {author}
  {\bibfnamefont {J.}~\bibnamefont {Millen}},\ }\bibfield  {title} {\bibinfo
  {title} {Full rotational control of levitated silicon nanorods},\ }\href
  {https://doi.org/10.1364/optica.4.000356} {\bibfield  {journal} {\bibinfo
  {journal} {Optica}\ }\textbf {\bibinfo {volume} {4}},\ \bibinfo {pages} {356}
  (\bibinfo {year} {2017})}\BibitemShut {NoStop}%
\bibitem [{\citenamefont {Mari}\ and\ \citenamefont {Eisert}(2009)}]{Mari2009}%
  \BibitemOpen
  \bibfield  {author} {\bibinfo {author} {\bibfnamefont {A.}~\bibnamefont
  {Mari}}\ and\ \bibinfo {author} {\bibfnamefont {J.}~\bibnamefont {Eisert}},\
  }\bibfield  {title} {\bibinfo {title} {{Gently Modulating Optomechanical
  Systems}},\ }\href {https://doi.org/10.1103/PhysRevLett.103.213603}
  {\bibfield  {journal} {\bibinfo  {journal} {Physical Review Letters}\
  }\textbf {\bibinfo {volume} {103}},\ \bibinfo {pages} {213603} (\bibinfo
  {year} {2009})}\BibitemShut {NoStop}%
\bibitem [{\citenamefont {Liao}\ and\ \citenamefont {Law}(2011)}]{Liao2011}%
  \BibitemOpen
  \bibfield  {author} {\bibinfo {author} {\bibfnamefont {J.-Q.}\ \bibnamefont
  {Liao}}\ and\ \bibinfo {author} {\bibfnamefont {C.~K.}\ \bibnamefont {Law}},\
  }\bibfield  {title} {\bibinfo {title} {Parametric generation of quadrature
  squeezing of mirrors in cavity optomechanics},\ }\href
  {https://doi.org/10.1103/PhysRevA.83.033820} {\bibfield  {journal} {\bibinfo
  {journal} {Physical Review A}\ }\textbf {\bibinfo {volume} {83}},\ \bibinfo
  {pages} {033820} (\bibinfo {year} {2011})}\BibitemShut {NoStop}%
\bibitem [{\citenamefont {Chowdhury}\ \emph {et~al.}(2019)\citenamefont
  {Chowdhury}, \citenamefont {Vezio}, \citenamefont {Bonaldi}, \citenamefont
  {Borrielli}, \citenamefont {Marino}, \citenamefont {Morana}, \citenamefont
  {Prodi}, \citenamefont {Sarro}, \citenamefont {Serra},\ and\ \citenamefont
  {Marin}}]{Chowdhury2019}%
  \BibitemOpen
  \bibfield  {author} {\bibinfo {author} {\bibfnamefont {A.}~\bibnamefont
  {Chowdhury}}, \bibinfo {author} {\bibfnamefont {P.}~\bibnamefont {Vezio}},
  \bibinfo {author} {\bibfnamefont {M.}~\bibnamefont {Bonaldi}}, \bibinfo
  {author} {\bibfnamefont {A.}~\bibnamefont {Borrielli}}, \bibinfo {author}
  {\bibfnamefont {F.}~\bibnamefont {Marino}}, \bibinfo {author} {\bibfnamefont
  {B.}~\bibnamefont {Morana}}, \bibinfo {author} {\bibfnamefont {G.~A.}\
  \bibnamefont {Prodi}}, \bibinfo {author} {\bibfnamefont {P.~M.}\ \bibnamefont
  {Sarro}}, \bibinfo {author} {\bibfnamefont {E.}~\bibnamefont {Serra}},\ and\
  \bibinfo {author} {\bibfnamefont {F.}~\bibnamefont {Marin}},\ }\bibfield
  {title} {\bibinfo {title} {Quantum signature of a squeezed mechanical
  oscillator},\ }\href {https://arxiv.org/abs/1907.05148} {\  (\bibinfo {year}
  {2019})},\ \Eprint {https://arxiv.org/abs/arXiv:1907.05148}
  {arXiv:1907.05148} \BibitemShut {NoStop}%
\bibitem [{\citenamefont {Kronwald}\ \emph {et~al.}(2013)\citenamefont
  {Kronwald}, \citenamefont {Marquardt},\ and\ \citenamefont
  {Clerk}}]{Kronwald2013}%
  \BibitemOpen
  \bibfield  {author} {\bibinfo {author} {\bibfnamefont {A.}~\bibnamefont
  {Kronwald}}, \bibinfo {author} {\bibfnamefont {F.}~\bibnamefont
  {Marquardt}},\ and\ \bibinfo {author} {\bibfnamefont {A.~A.}\ \bibnamefont
  {Clerk}},\ }\bibfield  {title} {\bibinfo {title} {{Arbitrarily large
  steady-state bosonic squeezing via dissipation}},\ }\href
  {https://doi.org/10.1103/PhysRevA.88.063833} {\bibfield  {journal} {\bibinfo
  {journal} {Physical Review A}\ }\textbf {\bibinfo {volume} {88}},\ \bibinfo
  {pages} {063833} (\bibinfo {year} {2013})}\BibitemShut {NoStop}%
\bibitem [{\citenamefont {Pirkkalainen}\ \emph {et~al.}(2015)\citenamefont
  {Pirkkalainen}, \citenamefont {Damsk{\"{a}}gg}, \citenamefont {Brandt},
  \citenamefont {Massel},\ and\ \citenamefont
  {Sillanp{\"{a}}{\"{a}}}}]{Pirkkalainen2015}%
  \BibitemOpen
  \bibfield  {author} {\bibinfo {author} {\bibfnamefont {J.-M.}\ \bibnamefont
  {Pirkkalainen}}, \bibinfo {author} {\bibfnamefont {E.}~\bibnamefont
  {Damsk{\"{a}}gg}}, \bibinfo {author} {\bibfnamefont {M.}~\bibnamefont
  {Brandt}}, \bibinfo {author} {\bibfnamefont {F.}~\bibnamefont {Massel}},\
  and\ \bibinfo {author} {\bibfnamefont {M.~A.}\ \bibnamefont
  {Sillanp{\"{a}}{\"{a}}}},\ }\bibfield  {title} {\bibinfo {title} {{Squeezing
  of Quantum Noise of Motion in a Micromechanical Resonator}},\ }\href
  {https://doi.org/10.1103/PhysRevLett.115.243601} {\bibfield  {journal}
  {\bibinfo  {journal} {Physical Review Letters}\ }\textbf {\bibinfo {volume}
  {115}},\ \bibinfo {pages} {243601} (\bibinfo {year} {2015})}\BibitemShut
  {NoStop}%
\bibitem [{\citenamefont {Wollman}\ \emph {et~al.}(2015)\citenamefont
  {Wollman}, \citenamefont {Lei}, \citenamefont {Weinstein}, \citenamefont
  {Suh}, \citenamefont {Kronwald}, \citenamefont {Marquardt}, \citenamefont
  {Clerk},\ and\ \citenamefont {Schwab}}]{Wollman2015}%
  \BibitemOpen
  \bibfield  {author} {\bibinfo {author} {\bibfnamefont {E.~E.}\ \bibnamefont
  {Wollman}}, \bibinfo {author} {\bibfnamefont {C.~U.}\ \bibnamefont {Lei}},
  \bibinfo {author} {\bibfnamefont {A.~J.}\ \bibnamefont {Weinstein}}, \bibinfo
  {author} {\bibfnamefont {J.}~\bibnamefont {Suh}}, \bibinfo {author}
  {\bibfnamefont {A.}~\bibnamefont {Kronwald}}, \bibinfo {author}
  {\bibfnamefont {F.}~\bibnamefont {Marquardt}}, \bibinfo {author}
  {\bibfnamefont {A.~A.}\ \bibnamefont {Clerk}},\ and\ \bibinfo {author}
  {\bibfnamefont {K.~C.}\ \bibnamefont {Schwab}},\ }\bibfield  {title}
  {\bibinfo {title} {{Quantum squeezing of motion in a mechanical resonator}},\
  }\href {https://doi.org/10.1126/science.aac5138} {\bibfield  {journal}
  {\bibinfo  {journal} {Science}\ }\textbf {\bibinfo {volume} {349}},\ \bibinfo
  {pages} {952} (\bibinfo {year} {2015})}\BibitemShut {NoStop}%
\bibitem [{\citenamefont {Lei}\ \emph {et~al.}(2016)\citenamefont {Lei},
  \citenamefont {Weinstein}, \citenamefont {Suh}, \citenamefont {Wollman},
  \citenamefont {Kronwald}, \citenamefont {Marquardt}, \citenamefont {Clerk},\
  and\ \citenamefont {Schwab}}]{Lei2016}%
  \BibitemOpen
  \bibfield  {author} {\bibinfo {author} {\bibfnamefont {C.~U.}\ \bibnamefont
  {Lei}}, \bibinfo {author} {\bibfnamefont {A.~J.}\ \bibnamefont {Weinstein}},
  \bibinfo {author} {\bibfnamefont {J.}~\bibnamefont {Suh}}, \bibinfo {author}
  {\bibfnamefont {E.~E.}\ \bibnamefont {Wollman}}, \bibinfo {author}
  {\bibfnamefont {A.}~\bibnamefont {Kronwald}}, \bibinfo {author}
  {\bibfnamefont {F.}~\bibnamefont {Marquardt}}, \bibinfo {author}
  {\bibfnamefont {A.~A.}\ \bibnamefont {Clerk}},\ and\ \bibinfo {author}
  {\bibfnamefont {K.~C.}\ \bibnamefont {Schwab}},\ }\bibfield  {title}
  {\bibinfo {title} {{Quantum Nondemolition Measurement of a Quantum Squeezed
  State Beyond the 3 dB Limit}},\ }\href
  {https://doi.org/10.1103/PhysRevLett.117.100801} {\bibfield  {journal}
  {\bibinfo  {journal} {Physical Review Letters}\ }\textbf {\bibinfo {volume}
  {117}},\ \bibinfo {pages} {100801} (\bibinfo {year} {2016})}\BibitemShut
  {NoStop}%
\bibitem [{\citenamefont {Clerk}\ \emph {et~al.}(2008)\citenamefont {Clerk},
  \citenamefont {Marquardt},\ and\ \citenamefont {Jacobs}}]{Clerk2008}%
  \BibitemOpen
  \bibfield  {author} {\bibinfo {author} {\bibfnamefont {A.~A.}\ \bibnamefont
  {Clerk}}, \bibinfo {author} {\bibfnamefont {F.}~\bibnamefont {Marquardt}},\
  and\ \bibinfo {author} {\bibfnamefont {K.}~\bibnamefont {Jacobs}},\
  }\bibfield  {title} {\bibinfo {title} {{Back-action evasion and squeezing of
  a mechanical resonator using a cavity detector}},\ }\href
  {https://doi.org/10.1088/1367-2630/10/9/095010} {\bibfield  {journal}
  {\bibinfo  {journal} {New Journal of Physics}\ }\textbf {\bibinfo {volume}
  {10}},\ \bibinfo {pages} {095010} (\bibinfo {year} {2008})}\BibitemShut
  {NoStop}%
\bibitem [{\citenamefont {Genoni}\ \emph {et~al.}(2015)\citenamefont {Genoni},
  \citenamefont {Zhang}, \citenamefont {Millen}, \citenamefont {Barker},\ and\
  \citenamefont {Serafini}}]{Genoni2015}%
  \BibitemOpen
  \bibfield  {author} {\bibinfo {author} {\bibfnamefont {M.~G.}\ \bibnamefont
  {Genoni}}, \bibinfo {author} {\bibfnamefont {J.}~\bibnamefont {Zhang}},
  \bibinfo {author} {\bibfnamefont {J.}~\bibnamefont {Millen}}, \bibinfo
  {author} {\bibfnamefont {P.~F.}\ \bibnamefont {Barker}},\ and\ \bibinfo
  {author} {\bibfnamefont {A.}~\bibnamefont {Serafini}},\ }\bibfield  {title}
  {\bibinfo {title} {Quantum cooling and squeezing of a levitating nanosphere
  via time-continuous measurements},\ }\href
  {https://doi.org/10.1088/1367-2630/17/7/073019} {\bibfield  {journal}
  {\bibinfo  {journal} {New Journal of Physics}\ }\textbf {\bibinfo {volume}
  {17}},\ \bibinfo {pages} {073019} (\bibinfo {year} {2015})}\BibitemShut
  {NoStop}%
\bibitem [{\citenamefont {Rakhubovsky}\ \emph {et~al.}(2019)\citenamefont
  {Rakhubovsky}, \citenamefont {Moore},\ and\ \citenamefont
  {Filip}}]{Rakhubovsky2019}%
  \BibitemOpen
  \bibfield  {author} {\bibinfo {author} {\bibfnamefont {A.~A.}\ \bibnamefont
  {Rakhubovsky}}, \bibinfo {author} {\bibfnamefont {D.~W.}\ \bibnamefont
  {Moore}},\ and\ \bibinfo {author} {\bibfnamefont {R.}~\bibnamefont {Filip}},\
  }\bibfield  {title} {\bibinfo {title} {Nonclassical states of levitated
  macroscopic objects beyond the ground state},\ }\href
  {https://doi.org/10.1088/2058-9565/ab043d} {\bibfield  {journal} {\bibinfo
  {journal} {Quantum Science and Technology}\ }\textbf {\bibinfo {volume}
  {4}},\ \bibinfo {pages} {024006} (\bibinfo {year} {2019})}\BibitemShut
  {NoStop}%
\bibitem [{\citenamefont {Nunnenkamp}\ \emph {et~al.}(2010)\citenamefont
  {Nunnenkamp}, \citenamefont {B{\o}rkje}, \citenamefont {Harris},\ and\
  \citenamefont {Girvin}}]{Nunnenkamp2010}%
  \BibitemOpen
  \bibfield  {author} {\bibinfo {author} {\bibfnamefont {A.}~\bibnamefont
  {Nunnenkamp}}, \bibinfo {author} {\bibfnamefont {K.}~\bibnamefont
  {B{\o}rkje}}, \bibinfo {author} {\bibfnamefont {J.~G.~E.}\ \bibnamefont
  {Harris}},\ and\ \bibinfo {author} {\bibfnamefont {S.~M.}\ \bibnamefont
  {Girvin}},\ }\bibfield  {title} {\bibinfo {title} {Cooling and squeezing via
  quadratic optomechanical coupling},\ }\href
  {https://doi.org/10.1103/physreva.82.021806} {\bibfield  {journal} {\bibinfo
  {journal} {Physical Review A}\ }\textbf {\bibinfo {volume} {82}},\ \bibinfo
  {pages} {021806} (\bibinfo {year} {2010})}\BibitemShut {NoStop}%
\bibitem [{\citenamefont {Asjad}\ \emph {et~al.}(2014)\citenamefont {Asjad},
  \citenamefont {Agarwal}, \citenamefont {Kim}, \citenamefont {Tombesi},
  \citenamefont {Giuseppe},\ and\ \citenamefont {Vitali}}]{Asjad2014}%
  \BibitemOpen
  \bibfield  {author} {\bibinfo {author} {\bibfnamefont {M.}~\bibnamefont
  {Asjad}}, \bibinfo {author} {\bibfnamefont {G.~S.}\ \bibnamefont {Agarwal}},
  \bibinfo {author} {\bibfnamefont {M.~S.}\ \bibnamefont {Kim}}, \bibinfo
  {author} {\bibfnamefont {P.}~\bibnamefont {Tombesi}}, \bibinfo {author}
  {\bibfnamefont {G.~D.}\ \bibnamefont {Giuseppe}},\ and\ \bibinfo {author}
  {\bibfnamefont {D.}~\bibnamefont {Vitali}},\ }\bibfield  {title} {\bibinfo
  {title} {Robust stationary mechanical squeezing in a kicked quadratic
  optomechanical system},\ }\href {https://doi.org/10.1103/physreva.89.023849}
  {\bibfield  {journal} {\bibinfo  {journal} {Physical Review A}\ }\textbf
  {\bibinfo {volume} {89}},\ \bibinfo {pages} {023849} (\bibinfo {year}
  {2014})}\BibitemShut {NoStop}%
\bibitem [{\citenamefont {L\"{u}}\ \emph {et~al.}(2015)\citenamefont {L\"{u}},
  \citenamefont {Liao}, \citenamefont {Tian},\ and\ \citenamefont
  {Nori}}]{Lu2015}%
  \BibitemOpen
  \bibfield  {author} {\bibinfo {author} {\bibfnamefont {X.-Y.}\ \bibnamefont
  {L\"{u}}}, \bibinfo {author} {\bibfnamefont {J.-Q.}\ \bibnamefont {Liao}},
  \bibinfo {author} {\bibfnamefont {L.}~\bibnamefont {Tian}},\ and\ \bibinfo
  {author} {\bibfnamefont {F.}~\bibnamefont {Nori}},\ }\bibfield  {title}
  {\bibinfo {title} {Steady-state mechanical squeezing in an optomechanical
  system via duffing nonlinearity},\ }\href
  {https://doi.org/10.1103/physreva.91.013834} {\bibfield  {journal} {\bibinfo
  {journal} {Physical Review A}\ }\textbf {\bibinfo {volume} {91}},\ \bibinfo
  {pages} {013834} (\bibinfo {year} {2015})}\BibitemShut {NoStop}%
\bibitem [{\citenamefont {Vuleti\'{c}}\ \emph {et~al.}(2001)\citenamefont
  {Vuleti\'{c}}, \citenamefont {Chan},\ and\ \citenamefont
  {Black}}]{Vuletic2001}%
  \BibitemOpen
  \bibfield  {author} {\bibinfo {author} {\bibfnamefont {V.}~\bibnamefont
  {Vuleti\'{c}}}, \bibinfo {author} {\bibfnamefont {H.~W.}\ \bibnamefont
  {Chan}},\ and\ \bibinfo {author} {\bibfnamefont {A.~T.}\ \bibnamefont
  {Black}},\ }\bibfield  {title} {\bibinfo {title} {Three-dimensional cavity
  {Doppler} cooling and cavity sideband cooling by coherent scattering},\
  }\href {https://doi.org/10.1103/PhysRevA.64.033405} {\bibfield  {journal}
  {\bibinfo  {journal} {Physical Review A}\ }\textbf {\bibinfo {volume} {64}},\
  \bibinfo {pages} {033405} (\bibinfo {year} {2001})}\BibitemShut {NoStop}%
\bibitem [{\citenamefont {Gonzalez-Ballestero}\ \emph
  {et~al.}(2019)\citenamefont {Gonzalez-Ballestero}, \citenamefont {Maurer},
  \citenamefont {Windey}, \citenamefont {Novotny}, \citenamefont {Reimann},\
  and\ \citenamefont {Romero-Isart}}]{Gonzalez-Ballestero2019}%
  \BibitemOpen
  \bibfield  {author} {\bibinfo {author} {\bibfnamefont {C.}~\bibnamefont
  {Gonzalez-Ballestero}}, \bibinfo {author} {\bibfnamefont {P.}~\bibnamefont
  {Maurer}}, \bibinfo {author} {\bibfnamefont {D.}~\bibnamefont {Windey}},
  \bibinfo {author} {\bibfnamefont {L.}~\bibnamefont {Novotny}}, \bibinfo
  {author} {\bibfnamefont {R.}~\bibnamefont {Reimann}},\ and\ \bibinfo {author}
  {\bibfnamefont {O.}~\bibnamefont {Romero-Isart}},\ }\bibfield  {title}
  {\bibinfo {title} {Theory for cavity cooling of levitated nanoparticles via
  coherent scattering: Master equation approach},\ }\href
  {https://doi.org/10.1103/PhysRevA.100.013805} {\bibfield  {journal} {\bibinfo
   {journal} {Physical Review A}\ }\textbf {\bibinfo {volume} {100}},\ \bibinfo
  {pages} {013805} (\bibinfo {year} {2019})}\BibitemShut {NoStop}%
\bibitem [{\citenamefont {Leibrandt}\ \emph {et~al.}(2009)\citenamefont
  {Leibrandt}, \citenamefont {Labaziewicz}, \citenamefont
  {Vuleti\ifmmode~\acute{c}\else \'{c}\fi{}},\ and\ \citenamefont
  {Chuang}}]{Leibrandt2009}%
  \BibitemOpen
  \bibfield  {author} {\bibinfo {author} {\bibfnamefont {D.~R.}\ \bibnamefont
  {Leibrandt}}, \bibinfo {author} {\bibfnamefont {J.}~\bibnamefont
  {Labaziewicz}}, \bibinfo {author} {\bibfnamefont {V.}~\bibnamefont
  {Vuleti\ifmmode~\acute{c}\else \'{c}\fi{}}},\ and\ \bibinfo {author}
  {\bibfnamefont {I.~L.}\ \bibnamefont {Chuang}},\ }\bibfield  {title}
  {\bibinfo {title} {{Cavity Sideband Cooling of a Single Trapped Ion}},\
  }\href {https://doi.org/10.1103/PhysRevLett.103.103001} {\bibfield  {journal}
  {\bibinfo  {journal} {Physical Review Letters}\ }\textbf {\bibinfo {volume}
  {103}},\ \bibinfo {pages} {103001} (\bibinfo {year} {2009})}\BibitemShut
  {NoStop}%
\bibitem [{\citenamefont {Deli{\'c}}\ \emph
  {et~al.}(2019{\natexlab{a}})\citenamefont {Deli{\'c}}, \citenamefont
  {Reisenbauer}, \citenamefont {Grass}, \citenamefont {Kiesel}, \citenamefont
  {Vuleti{\'c}},\ and\ \citenamefont {Aspelmeyer}}]{Delic2019}%
  \BibitemOpen
  \bibfield  {author} {\bibinfo {author} {\bibfnamefont {U.}~\bibnamefont
  {Deli{\'c}}}, \bibinfo {author} {\bibfnamefont {M.}~\bibnamefont
  {Reisenbauer}}, \bibinfo {author} {\bibfnamefont {D.}~\bibnamefont {Grass}},
  \bibinfo {author} {\bibfnamefont {N.}~\bibnamefont {Kiesel}}, \bibinfo
  {author} {\bibfnamefont {V.}~\bibnamefont {Vuleti{\'c}}},\ and\ \bibinfo
  {author} {\bibfnamefont {M.}~\bibnamefont {Aspelmeyer}},\ }\bibfield  {title}
  {\bibinfo {title} {{Cavity Cooling of a Levitated Nanosphere by Coherent
  Scattering}},\ }\href {https://doi.org/10.1103/PhysRevLett.122.123602}
  {\bibfield  {journal} {\bibinfo  {journal} {Physical Review Letters}\
  }\textbf {\bibinfo {volume} {122}},\ \bibinfo {pages} {123602} (\bibinfo
  {year} {2019}{\natexlab{a}})}\BibitemShut {NoStop}%
\bibitem [{\citenamefont {Windey}\ \emph {et~al.}(2019)\citenamefont {Windey},
  \citenamefont {Gonzalez-Ballestero}, \citenamefont {Maurer}, \citenamefont
  {Novotny}, \citenamefont {Romero-Isart},\ and\ \citenamefont
  {Reimann}}]{Windey2019}%
  \BibitemOpen
  \bibfield  {author} {\bibinfo {author} {\bibfnamefont {D.}~\bibnamefont
  {Windey}}, \bibinfo {author} {\bibfnamefont {C.}~\bibnamefont
  {Gonzalez-Ballestero}}, \bibinfo {author} {\bibfnamefont {P.}~\bibnamefont
  {Maurer}}, \bibinfo {author} {\bibfnamefont {L.}~\bibnamefont {Novotny}},
  \bibinfo {author} {\bibfnamefont {O.}~\bibnamefont {Romero-Isart}},\ and\
  \bibinfo {author} {\bibfnamefont {R.}~\bibnamefont {Reimann}},\ }\bibfield
  {title} {\bibinfo {title} {{Cavity-Based 3D Cooling of a Levitated
  Nanoparticle via Coherent Scattering}},\ }\href
  {https://doi.org/10.1103/PhysRevLett.122.123601} {\bibfield  {journal}
  {\bibinfo  {journal} {Physical Review Letters}\ }\textbf {\bibinfo {volume}
  {122}},\ \bibinfo {pages} {123601} (\bibinfo {year} {2019})}\BibitemShut
  {NoStop}%
\bibitem [{\citenamefont {Deli{\'c}}\ \emph
  {et~al.}(2019{\natexlab{b}})\citenamefont {Deli{\'c}}, \citenamefont
  {Reisenbauer}, \citenamefont {Dare}, \citenamefont {Grass}, \citenamefont
  {Vuleti{\'c}}, \citenamefont {Kiesel},\ and\ \citenamefont
  {Aspelmeyer}}]{Delic2019b}%
  \BibitemOpen
  \bibfield  {author} {\bibinfo {author} {\bibfnamefont {U.}~\bibnamefont
  {Deli{\'c}}}, \bibinfo {author} {\bibfnamefont {M.}~\bibnamefont
  {Reisenbauer}}, \bibinfo {author} {\bibfnamefont {K.}~\bibnamefont {Dare}},
  \bibinfo {author} {\bibfnamefont {D.}~\bibnamefont {Grass}}, \bibinfo
  {author} {\bibfnamefont {V.}~\bibnamefont {Vuleti{\'c}}}, \bibinfo {author}
  {\bibfnamefont {N.}~\bibnamefont {Kiesel}},\ and\ \bibinfo {author}
  {\bibfnamefont {M.}~\bibnamefont {Aspelmeyer}},\ }\bibfield  {title}
  {\bibinfo {title} {Motional quantum ground state of a levitated nanoparticle
  from room temperature},\ }\href {https://arxiv.org/abs/1911.04406} {\
  (\bibinfo {year} {2019}{\natexlab{b}})},\ \Eprint
  {https://arxiv.org/abs/arXiv:1911.04406} {arXiv:1911.04406} \BibitemShut
  {NoStop}%
\bibitem [{\citenamefont {Hu}\ \emph {et~al.}(2018)\citenamefont {Hu},
  \citenamefont {Yang}, \citenamefont {Wu}, \citenamefont {Li},\ and\
  \citenamefont {Zheng}}]{Hu2018}%
  \BibitemOpen
  \bibfield  {author} {\bibinfo {author} {\bibfnamefont {C.-S.}\ \bibnamefont
  {Hu}}, \bibinfo {author} {\bibfnamefont {Z.-B.}\ \bibnamefont {Yang}},
  \bibinfo {author} {\bibfnamefont {H.}~\bibnamefont {Wu}}, \bibinfo {author}
  {\bibfnamefont {Y.}~\bibnamefont {Li}},\ and\ \bibinfo {author}
  {\bibfnamefont {S.-B.}\ \bibnamefont {Zheng}},\ }\bibfield  {title} {\bibinfo
  {title} {Twofold mechanical squeezing in a cavity optomechanical system},\
  }\href {https://doi.org/10.1103/PhysRevA.98.023807} {\bibfield  {journal}
  {\bibinfo  {journal} {Physical Review A}\ }\textbf {\bibinfo {volume} {98}},\
  \bibinfo {pages} {023807} (\bibinfo {year} {2018})}\BibitemShut {NoStop}%
\bibitem [{\citenamefont {Asjad}\ \emph {et~al.}(2016)\citenamefont {Asjad},
  \citenamefont {Zippilli},\ and\ \citenamefont {Vitali}}]{Asjad2016}%
  \BibitemOpen
  \bibfield  {author} {\bibinfo {author} {\bibfnamefont {M.}~\bibnamefont
  {Asjad}}, \bibinfo {author} {\bibfnamefont {S.}~\bibnamefont {Zippilli}},\
  and\ \bibinfo {author} {\bibfnamefont {D.}~\bibnamefont {Vitali}},\
  }\bibfield  {title} {\bibinfo {title} {{Suppression of Stokes scattering and
  improved optomechanical cooling with squeezed light}},\ }\href
  {https://doi.org/10.1103/PhysRevA.94.051801} {\bibfield  {journal} {\bibinfo
  {journal} {Physical Review A}\ }\textbf {\bibinfo {volume} {94}},\ \bibinfo
  {pages} {051801} (\bibinfo {year} {2016})}\BibitemShut {NoStop}%
\bibitem [{\citenamefont {Clark}\ \emph {et~al.}(2017)\citenamefont {Clark},
  \citenamefont {Lecocq}, \citenamefont {Simmonds}, \citenamefont {Aumentado},\
  and\ \citenamefont {Teufel}}]{Clark2017}%
  \BibitemOpen
  \bibfield  {author} {\bibinfo {author} {\bibfnamefont {J.~B.}\ \bibnamefont
  {Clark}}, \bibinfo {author} {\bibfnamefont {F.}~\bibnamefont {Lecocq}},
  \bibinfo {author} {\bibfnamefont {R.~W.}\ \bibnamefont {Simmonds}}, \bibinfo
  {author} {\bibfnamefont {J.}~\bibnamefont {Aumentado}},\ and\ \bibinfo
  {author} {\bibfnamefont {J.~D.}\ \bibnamefont {Teufel}},\ }\bibfield  {title}
  {\bibinfo {title} {{Sideband Cooling Beyond the Quantum Limit with Squeezed
  Light}},\ }\href {https://doi.org/10.1038/nature20604} {\bibfield  {journal}
  {\bibinfo  {journal} {Nature}\ }\textbf {\bibinfo {volume} {541}},\ \bibinfo
  {pages} {191} (\bibinfo {year} {2017})}\BibitemShut {NoStop}%
\bibitem [{\citenamefont {Pietik\"{a}inen}\ \emph {et~al.}()\citenamefont
  {Pietik\"{a}inen}, \citenamefont {\v{C}ernot\'ik},\ and\ \citenamefont
  {Filip}}]{Pietikainen2019}%
  \BibitemOpen
  \bibfield  {author} {\bibinfo {author} {\bibfnamefont {I.}~\bibnamefont
  {Pietik\"{a}inen}}, \bibinfo {author} {\bibfnamefont {O.}~\bibnamefont
  {\v{C}ernot\'ik}},\ and\ \bibinfo {author} {\bibfnamefont {R.}~\bibnamefont
  {Filip}},\ }\href@noop {} {}\bibinfo {note} {In preparation}\BibitemShut
  {NoStop}%
\bibitem [{\citenamefont {Pontin}\ \emph {et~al.}(2016)\citenamefont {Pontin},
  \citenamefont {Bonaldi}, \citenamefont {Borrielli}, \citenamefont {Marconi},
  \citenamefont {Marino}, \citenamefont {Pandraud}, \citenamefont {Prodi},
  \citenamefont {Sarro}, \citenamefont {Serra},\ and\ \citenamefont
  {Marin}}]{Pontin2016}%
  \BibitemOpen
  \bibfield  {author} {\bibinfo {author} {\bibfnamefont {A.}~\bibnamefont
  {Pontin}}, \bibinfo {author} {\bibfnamefont {M.}~\bibnamefont {Bonaldi}},
  \bibinfo {author} {\bibfnamefont {A.}~\bibnamefont {Borrielli}}, \bibinfo
  {author} {\bibfnamefont {L.}~\bibnamefont {Marconi}}, \bibinfo {author}
  {\bibfnamefont {F.}~\bibnamefont {Marino}}, \bibinfo {author} {\bibfnamefont
  {G.}~\bibnamefont {Pandraud}}, \bibinfo {author} {\bibfnamefont {G.~A.}\
  \bibnamefont {Prodi}}, \bibinfo {author} {\bibfnamefont {P.~M.}\ \bibnamefont
  {Sarro}}, \bibinfo {author} {\bibfnamefont {E.}~\bibnamefont {Serra}},\ and\
  \bibinfo {author} {\bibfnamefont {F.}~\bibnamefont {Marin}},\ }\bibfield
  {title} {\bibinfo {title} {{Dynamical Two-Mode Squeezing of Thermal
  Fluctuations in a Cavity Optomechanical System}},\ }\href
  {https://doi.org/10.1103/PhysRevLett.116.103601} {\bibfield  {journal}
  {\bibinfo  {journal} {Physical Review Letters}\ }\textbf {\bibinfo {volume}
  {116}},\ \bibinfo {pages} {103601} (\bibinfo {year} {2016})}\BibitemShut
  {NoStop}%
\bibitem [{\citenamefont {Ockeloen-Korppi}\ \emph
  {et~al.}(2018{\natexlab{b}})\citenamefont {Ockeloen-Korppi}, \citenamefont
  {Damskagg}, \citenamefont {Pirkkalainen}, \citenamefont {Clerk},
  \citenamefont {Massel}, \citenamefont {Woolley},\ and\ \citenamefont
  {Sillanpaa}}]{Ockeloen-Korppi2018}%
  \BibitemOpen
  \bibfield  {author} {\bibinfo {author} {\bibfnamefont {C.~F.}\ \bibnamefont
  {Ockeloen-Korppi}}, \bibinfo {author} {\bibfnamefont {E.}~\bibnamefont
  {Damskagg}}, \bibinfo {author} {\bibfnamefont {J.-M.}\ \bibnamefont
  {Pirkkalainen}}, \bibinfo {author} {\bibfnamefont {A.~A.}\ \bibnamefont
  {Clerk}}, \bibinfo {author} {\bibfnamefont {F.}~\bibnamefont {Massel}},
  \bibinfo {author} {\bibfnamefont {M.~J.}\ \bibnamefont {Woolley}},\ and\
  \bibinfo {author} {\bibfnamefont {M.~A.}\ \bibnamefont {Sillanpaa}},\
  }\bibfield  {title} {\bibinfo {title} {Stabilized entanglement of massive
  mechanical oscillators},\ }\href {https://doi.org/10.1038/s41586-018-0038-x}
  {\bibfield  {journal} {\bibinfo  {journal} {Nature}\ }\textbf {\bibinfo
  {volume} {556}},\ \bibinfo {pages} {478} (\bibinfo {year}
  {2018}{\natexlab{b}})}\BibitemShut {NoStop}%
\bibitem [{\citenamefont {Neumeier}\ \emph {et~al.}(2018)\citenamefont
  {Neumeier}, \citenamefont {Northup},\ and\ \citenamefont
  {Chang}}]{Neumeier2018}%
  \BibitemOpen
  \bibfield  {author} {\bibinfo {author} {\bibfnamefont {L.}~\bibnamefont
  {Neumeier}}, \bibinfo {author} {\bibfnamefont {T.~E.}\ \bibnamefont
  {Northup}},\ and\ \bibinfo {author} {\bibfnamefont {D.~E.}\ \bibnamefont
  {Chang}},\ }\bibfield  {title} {\bibinfo {title} {Reaching the optomechanical
  strong coupling regime with a single atom in a cavity},\ }\href
  {https://doi.org/10.1103/PhysRevA.97.063857} {\bibfield  {journal} {\bibinfo
  {journal} {Physical Review A}\ }\textbf {\bibinfo {volume} {97}},\ \bibinfo
  {pages} {063857} (\bibinfo {year} {2018})}\BibitemShut {NoStop}%
\bibitem [{\citenamefont {Deli{\'c}}\ \emph {et~al.}()\citenamefont
  {Deli{\'c}}, \citenamefont {Grass}, \citenamefont {Reisenbauer},
  \citenamefont {Damm}, \citenamefont {Weitz}, \citenamefont {Kiesel},\ and\
  \citenamefont {Aspelmeyer}}]{Delic2019a}%
  \BibitemOpen
  \bibfield  {author} {\bibinfo {author} {\bibfnamefont {U.}~\bibnamefont
  {Deli{\'c}}}, \bibinfo {author} {\bibfnamefont {D.}~\bibnamefont {Grass}},
  \bibinfo {author} {\bibfnamefont {M.}~\bibnamefont {Reisenbauer}}, \bibinfo
  {author} {\bibfnamefont {T.}~\bibnamefont {Damm}}, \bibinfo {author}
  {\bibfnamefont {M.}~\bibnamefont {Weitz}}, \bibinfo {author} {\bibfnamefont
  {N.}~\bibnamefont {Kiesel}},\ and\ \bibinfo {author} {\bibfnamefont
  {M.}~\bibnamefont {Aspelmeyer}},\ }\bibfield  {title} {\bibinfo {title}
  {Levitated cavity optomechanics in high vacuum},\ }\href
  {https://arxiv.org/abs/1902.06605} {\ }\Eprint
  {https://arxiv.org/abs/arXiv:1902.06605} {arXiv:1902.06605} \BibitemShut
  {NoStop}%
\bibitem [{\citenamefont {Rakhubovsky}\ and\ \citenamefont
  {Filip}()}]{Rakhubovsky2019a}%
  \BibitemOpen
  \bibfield  {author} {\bibinfo {author} {\bibfnamefont {A.~A.}\ \bibnamefont
  {Rakhubovsky}}\ and\ \bibinfo {author} {\bibfnamefont {R.}~\bibnamefont
  {Filip}},\ }\bibfield  {title} {\bibinfo {title} {Stroboscopic high-order
  nonlinearity in quantum optomechanics},\ }\href
  {https://arxiv.org/abs/1904.00773} {\ }\Eprint
  {https://arxiv.org/abs/arXiv:1904.00773} {arXiv:1904.00773} \BibitemShut
  {NoStop}%
\bibitem [{\citenamefont {Setter}\ \emph {et~al.}(2019)\citenamefont {Setter},
  \citenamefont {Vovrosh},\ and\ \citenamefont {Ulbricht}}]{Setter2019}%
  \BibitemOpen
  \bibfield  {author} {\bibinfo {author} {\bibfnamefont {A.}~\bibnamefont
  {Setter}}, \bibinfo {author} {\bibfnamefont {J.}~\bibnamefont {Vovrosh}},\
  and\ \bibinfo {author} {\bibfnamefont {H.}~\bibnamefont {Ulbricht}},\
  }\bibfield  {title} {\bibinfo {title} {Characterization of non-linearities
  through mechanical squeezing in levitated optomechanics},\ }\href
  {https://doi.org/10.1063/1.5116121} {\bibfield  {journal} {\bibinfo
  {journal} {Applied Physics Letters}\ }\textbf {\bibinfo {volume} {115}},\
  \bibinfo {pages} {153106} (\bibinfo {year} {2019})}\BibitemShut {NoStop}%
\bibitem [{\citenamefont {Riedinger}\ \emph {et~al.}(2016)\citenamefont
  {Riedinger}, \citenamefont {Hong}, \citenamefont {Norte}, \citenamefont
  {Slater}, \citenamefont {Shang}, \citenamefont {Krause}, \citenamefont
  {Anant}, \citenamefont {Aspelmeyer},\ and\ \citenamefont
  {Gr{\"{o}}blacher}}]{Riedinger2016}%
  \BibitemOpen
  \bibfield  {author} {\bibinfo {author} {\bibfnamefont {R.}~\bibnamefont
  {Riedinger}}, \bibinfo {author} {\bibfnamefont {S.}~\bibnamefont {Hong}},
  \bibinfo {author} {\bibfnamefont {R.~A.}\ \bibnamefont {Norte}}, \bibinfo
  {author} {\bibfnamefont {J.~A.}\ \bibnamefont {Slater}}, \bibinfo {author}
  {\bibfnamefont {J.}~\bibnamefont {Shang}}, \bibinfo {author} {\bibfnamefont
  {A.~G.}\ \bibnamefont {Krause}}, \bibinfo {author} {\bibfnamefont
  {V.}~\bibnamefont {Anant}}, \bibinfo {author} {\bibfnamefont
  {M.}~\bibnamefont {Aspelmeyer}},\ and\ \bibinfo {author} {\bibfnamefont
  {S.}~\bibnamefont {Gr{\"{o}}blacher}},\ }\bibfield  {title} {\bibinfo {title}
  {{Non-classical correlations between single photons and phonons from a
  mechanical oscillator}},\ }\href {https://doi.org/10.1038/nature16536}
  {\bibfield  {journal} {\bibinfo  {journal} {Nature}\ }\textbf {\bibinfo
  {volume} {530}},\ \bibinfo {pages} {313} (\bibinfo {year}
  {2016})}\BibitemShut {NoStop}%
\bibitem [{\citenamefont {Chu}\ \emph {et~al.}(2018)\citenamefont {Chu},
  \citenamefont {Kharel}, \citenamefont {Yoon}, \citenamefont {Frunzio},
  \citenamefont {Rakich},\ and\ \citenamefont {Schoelkopf}}]{Chu2018}%
  \BibitemOpen
  \bibfield  {author} {\bibinfo {author} {\bibfnamefont {Y.}~\bibnamefont
  {Chu}}, \bibinfo {author} {\bibfnamefont {P.}~\bibnamefont {Kharel}},
  \bibinfo {author} {\bibfnamefont {T.}~\bibnamefont {Yoon}}, \bibinfo {author}
  {\bibfnamefont {L.}~\bibnamefont {Frunzio}}, \bibinfo {author} {\bibfnamefont
  {P.~T.}\ \bibnamefont {Rakich}},\ and\ \bibinfo {author} {\bibfnamefont
  {R.~J.}\ \bibnamefont {Schoelkopf}},\ }\bibfield  {title} {\bibinfo {title}
  {Creation and control of multi-phonon fock states in a bulk acoustic-wave
  resonator},\ }\href {https://doi.org/10.1038/s41586-018-0717-7} {\bibfield
  {journal} {\bibinfo  {journal} {Nature}\ }\textbf {\bibinfo {volume} {563}},\
  \bibinfo {pages} {666} (\bibinfo {year} {2018})}\BibitemShut {NoStop}%
\bibitem [{\citenamefont {Satzinger}\ \emph {et~al.}(2018)\citenamefont
  {Satzinger}, \citenamefont {Zhong}, \citenamefont {Chang}, \citenamefont
  {Peairs}, \citenamefont {Bienfait}, \citenamefont {Chou}, \citenamefont
  {Cleland}, \citenamefont {Conner}, \citenamefont {Dumur}, \citenamefont
  {Grebel}, \citenamefont {Gutierrez}, \citenamefont {November}, \citenamefont
  {Povey}, \citenamefont {Whiteley}, \citenamefont {Awschalom}, \citenamefont
  {Schuster},\ and\ \citenamefont {Cleland}}]{Satzinger2018}%
  \BibitemOpen
  \bibfield  {author} {\bibinfo {author} {\bibfnamefont {K.~J.}\ \bibnamefont
  {Satzinger}}, \bibinfo {author} {\bibfnamefont {Y.~P.}\ \bibnamefont
  {Zhong}}, \bibinfo {author} {\bibfnamefont {H.-S.}\ \bibnamefont {Chang}},
  \bibinfo {author} {\bibfnamefont {G.~A.}\ \bibnamefont {Peairs}}, \bibinfo
  {author} {\bibfnamefont {A.}~\bibnamefont {Bienfait}}, \bibinfo {author}
  {\bibfnamefont {M.-H.}\ \bibnamefont {Chou}}, \bibinfo {author}
  {\bibfnamefont {A.~Y.}\ \bibnamefont {Cleland}}, \bibinfo {author}
  {\bibfnamefont {C.~R.}\ \bibnamefont {Conner}}, \bibinfo {author}
  {\bibfnamefont {E.}~\bibnamefont {Dumur}}, \bibinfo {author} {\bibfnamefont
  {J.}~\bibnamefont {Grebel}}, \bibinfo {author} {\bibfnamefont
  {I.}~\bibnamefont {Gutierrez}}, \bibinfo {author} {\bibfnamefont {B.~H.}\
  \bibnamefont {November}}, \bibinfo {author} {\bibfnamefont {R.~G.}\
  \bibnamefont {Povey}}, \bibinfo {author} {\bibfnamefont {S.~J.}\ \bibnamefont
  {Whiteley}}, \bibinfo {author} {\bibfnamefont {D.~D.}\ \bibnamefont
  {Awschalom}}, \bibinfo {author} {\bibfnamefont {D.~I.}\ \bibnamefont
  {Schuster}},\ and\ \bibinfo {author} {\bibfnamefont {A.~N.}\ \bibnamefont
  {Cleland}},\ }\bibfield  {title} {\bibinfo {title} {Quantum control of
  surface acoustic-wave phonons},\ }\href
  {https://doi.org/10.1038/s41586-018-0719-5} {\bibfield  {journal} {\bibinfo
  {journal} {Nature}\ }\textbf {\bibinfo {volume} {563}},\ \bibinfo {pages}
  {661} (\bibinfo {year} {2018})}\BibitemShut {NoStop}%
\bibitem [{\citenamefont {{\v{C}}ernot{\'{i}}k}\ \emph
  {et~al.}(2015)\citenamefont {{\v{C}}ernot{\'{i}}k}, \citenamefont
  {Vasilyev},\ and\ \citenamefont {Hammerer}}]{Cernotik2015}%
  \BibitemOpen
  \bibfield  {author} {\bibinfo {author} {\bibfnamefont {O.}~\bibnamefont
  {{\v{C}}ernot{\'{i}}k}}, \bibinfo {author} {\bibfnamefont {D.~V.}\
  \bibnamefont {Vasilyev}},\ and\ \bibinfo {author} {\bibfnamefont
  {K.}~\bibnamefont {Hammerer}},\ }\bibfield  {title} {\bibinfo {title}
  {{Adiabatic elimination of Gaussian subsystems from quantum dynamics under
  continuous measurement}},\ }\href
  {https://doi.org/10.1103/PhysRevA.92.012124} {\bibfield  {journal} {\bibinfo
  {journal} {Physical Review A}\ }\textbf {\bibinfo {volume} {92}},\ \bibinfo
  {pages} {012124} (\bibinfo {year} {2015})}\BibitemShut {NoStop}%
\end{thebibliography}
%

\end{document}